\documentclass[fleqn,usenatbib]{mnras}
\usepackage[T1]{fontenc}
\usepackage{graphicx}
\usepackage{amsmath}
\usepackage{amssymb}
\usepackage{newtxtext,newtxmath}
\usepackage[symbol]{footmisc}
\usepackage{textcase}
\usepackage{color}
\usepackage{soul}

\newcommand{\Reff}{\ensuremath{R_{\rm e}}}
\newcommand{\bn}{\ensuremath{b_n}}
\newcommand{\Rn}{\ensuremath{R^{1/n}}}
\newcommand{\rthree}{\ensuremath{r_{-3}}}

\title[$R_{\rm e}$, or not $R_{\rm e}$]{$R_{\rm e}$, or not $R_{\rm e}$: Developing $R_5\equiv R_{-2}$ as a scale radius for galaxy sizes, masses, and mass-to-light ratios}
\author[Graham]
{
Alister W.\ Graham$^1$\thanks{E-mail: AGraham@swin.edu.au}
\\
$^1$ Centre for Astrophysics and Supercomputing, Swinburne University of Technology, Hawthorn, VIC 3122, Australia
}

\date{Accepted XXX. Received YYY; in original form ZZZ}
\pubyear{2026}

\begin{document}
\label{firstpage}
\pagerange{\pageref{firstpage}--\pageref{lastpage}}
\maketitle

\begin{abstract}
The effective half-light radius $\text{$R_{\text{e}}$}$ marks an arbitrary 50 per cent
light boundary, and scaling relations involving such {\it effective} radii and their 
associated surface brightnesses, $\mu_{\text{e}}$, systematically (and undesirably) vary as the percentage changes.
Here, the projected radius $R_5\equiv R_{-2}$, where the logarithmic slope of the 
surface-brightness and intensity profiles equal $5.00\,\text{mag\,dex}^{-1}$ and $-2$, respectively, 
and where the luminosity per logarithmic radial interval is maximal, is 
developed as an alternative. 
It can be measured non-parametrically or with 
a parametrized fit.  For the S\'ersic $R^{1/n}$ family,  
$R_5=(2n/b_n)^n\,\text{$R_{\text{e}}$}$, with 
$R_5/\text{$R_{\text{e}}$}\rightarrow{\mathrm{e}}^{1/6}\approx1.181$ as $n\rightarrow\infty$. 
Reparameterizing the (now $b_n$-free) $R^{1/n}$ model in terms of the observable pair 
$(R_5,\mu_5)$ removes the non-linear $\text{$R_{\text{e}}$}$--$n$ coupling, 
and because the local slope is $5\,\text{mag\,dex}^{-1}$ at $R_5$, correlated measurement 
errors in $R_5$ and $\mu_5$ better cancel when deriving the inferred total magnitude 
($\propto\mu_5-5\log R_5$).
Additionally, an exact single-integral identity is provided to relate  
any projected light percentage to the percentage within a sphere of the same radius.
The scale radius $R_5$ is shown to be connected, through a weakly $n$-dependent factor, $g(n)$, to the
anisotropy-insensitive internal/spatial radius $r_{-3}$, yielding a refined $n$-dependent Wolf-type 
mass estimator $M_{-3}$ and spatial mass-to-light ratio $(M_{\mathrm{dyn}}/L)_{-3}$.  
Specifically, $M_{-3}\equiv M(<r_{-3})=3\,G^{-1}\langle\sigma_{\mathrm{los}}^2\rangle\,g(n)\,R_{-2}$, 
and for $n\gtrsim2$, $M_{-3}\approx4\,G^{-1}\langle\sigma_{\mathrm{los}}^2\rangle\,R_{-2}\approx4.72\,G^{-1}\langle\sigma_{\mathrm{los}}^2\rangle\,\text{$R_{\text{e}}$}$. 
Past half-light substitutions in dynamical mass estimators introduce systematic 
S\'ersic-dependent offsets of 12--18 per~cent in enclosed mass and offsets spanning $>20$ 
per~cent in the mass-to-light ratio within $r_{-3}$. 
Dynamical mass estimators appropriate for (dark matter)-free stellar systems are also provided,
and all relations are implemented in the companion Python package R5Tools.
\end{abstract}



\section{Introduction}
\label{Sec_Intro}

When \citet[][]{1963BAAA....6...41S} introduced the $R^{1/n}$ model as a generalization of the \citet[][]{1948AnAp...11..247D} $R^{1/4}$ model for describing the projected (as seen on the plane of the sky) radial distribution of light in galaxies, it had the following form
\begin{equation}
I(R) = 10^{-0.4 \mu_0} \cdot {\rm exp}\left[-n \left(\frac{R}{R_0}\right)^{1/n}\right], 
\label{eq:sersic_original}
\end{equation}
such that the intensity, $I$, declined with the projected (two-dimensions: 2D) radius, $R$.  
The parameter $n$ quantifies the curvature (a.k.a.\ shape or radial concentration) in the intensity profile (a.k.a.\ light profile). 
The central surface brightness $\mu_0 = -2.5\log I(0) \equiv -2.5\log I_0$, 
and the parameter $R_0$ marks the projected radius where the intensity has dropped to ${\rm e}^{-n}$ of its central value, which differs from a traditional scale-length, where the decline from the central value is a factor of ${\rm e}^1\approx 2.718$, as used in the expression from \citet{1988MNRAS.232..239D}. 

The $R^{1/4}$ model had been popular for many years and it instead used an `effective half light' scale radius, $R_{\rm e}$, denoting the 2D radius that defined a cylinder centered on the galaxy and enclosing half of the total light. \citet{1985LNP...232...53C, 1987IAUS..127...47C} cleverly incorporated $R_{\rm e}$ into the $R^{1/n}$ model, by re-expressing Equation~\ref{eq:sersic_original} as 
\begin{equation}
I(R)=I_{\rm e}\exp\left\{ -b_n\left[\left( \frac{R}{R_{\rm e}}\right) ^{1/n}
  -1\right]\right\},
\label{Eq_Ser}
\end{equation}
with $b_n \approx 1.9992n-0.3271$ for $0.5<n<10$ \citep{1989woga.conf..208C} ensuring that $R_{\rm e}$ encloses half of the $R^{1/n}$ model's total light.\footnote{For $n=0.5$, $b_n=\ln 2$, exactly.} 
%
%
This shift to $R_{\rm e}$ avoided the small scale-lengths associated with the steep inner nature of high-$n$ light profiles.  
Later works by \citet{1991A&A...249...99C}, \citet{1993MNRAS.265.1013C}, and \citet{1994MNRAS.271..523D} cemented the use of $R_{\rm e}$, and a review of the $R^{1/n}$ model and its associated expressions can be found in \citet{2005PASA...22..118G}.

The effective half-light radius, and the associated intensity $I_{\rm e}$, 
has undoubtedly proved highly useful. However, just as the $R^{1/4}$ model 
was based on an arbitrary exponent ($1/4$), the effective radius is based 
on an arbitrary percentage of light (50 per cent). 
As first explained in 
\citet{2003AJ....125.2936G} and developed further in 
\citet{2019PASA...36...35G}, $R_{\rm e}$ has also been highly misleading at times. 
\citet{2019PASA...36...35G} examined an extensive array of alternative percentages and scale radii and showed how the curved 
 (non-log-linear) nature of scaling relations involving either $R_{\rm e}$ or $\mu_{\rm e}$, the surface brightness at $R_{\rm e}$, led to somewhat misplaced claims of a dichotomy in the formation physics of early-type galaxies (ETGs).\footnote{
Observationally,
\citet{2019ApJ...872L..14M} and \citet{2019ApJ...872L..13M} 
likewise
demonstrated that altering the enclosed-light definition (e.g., using 20, 50
or 80 per cent light radii) significantly shifts scaling relations, alters
the apparent structural separation between quiescent and star-forming
galaxies, and connects differently to halo mass. The present work therefore
moves beyond any fixed-percentage definition and adopts a gradient-defined
scale.
}
The subsequent study of ultra-diffuse galaxies (UDGs) by 
\citet{2025PASA...42..155G} reinforced the point by demonstrating that UDGs
extend the ordinary--dwarf ETG sequence along a series of curved (and predicted) relations to fainter systems with larger sizes, while also noting that isophotal radii, although useful for higher-surface-brightness systems, become 
impractical for the low-surface-brightness UDG population. The present 
paper extends those works by replacing the arbitrary 
50-per-cent radius with a gradient-defined alternative, and by developing the corresponding structural equations and 
dynamical machinery to estimate masses and mass-to-light ratios. 

The scale radius, denoted $R_5$, is defined where the logarithmic slope of the surface brightness profile equals 
$5.00\,\mathrm{mag\,dex^{-1}}$, corresponding to
$\mathrm{d}\ln I/\mathrm{d}\ln R=-2$ (Section~\ref{Sec_log}). It was initially identified from where the linear slope of the surface brightness profile equals a rather unassuming value of $\approx 1.822$ (Section~\ref{Sec_linear}).  
Beyond its simple gradient-based definition, $R_5$ has a natural physical interpretation: it is the projected
radius at which the luminosity contribution per logarithmic interval
of radius, $\mathrm{d}L/\mathrm{d}\ln R\propto R^2I(R)$, reaches its maximum (Section~\ref{Sec_benefit_phys}). 
The $R_5$ radial scale is roughly 20 per cent larger than $R_{\rm e}$ and has several observational advantages noted in Section~\ref{Sec_benefit}.  
The associated surface brightness, $\mu_5$, provides a useful companion observable, with 
the $R_5,\mu_5$ parameterization removing the $R^{1/n}$ model's explicit
dependence on both $R_{\rm e}$ and the binding parameter $b_n$,
simplifying the analytical formulation of the profile and 
substantially reducing the non-linear coupling
among the structural parameters. Furthermore, 
correlated errors in $R_5$ and $\mu_5$ 
along the one-dimensional profile analytically compensate to first order
in the determination of the total magnitude, $m_{\rm tot} \propto \mu_5 - 5\log\, R_5$ (Section~\ref{Sec_stable}). 
Because $R_5$ can also be identified directly from the
observed surface-brightness gradient, a non-parametric procedure is
developed for estimating the concentration, S\'ersic index, and total
luminosity without performing a conventional S\'ersic fit (Section~\ref{Sec_Non-par}). 

Section~\ref{Sec_new_rel} reformulates the $R^{1/n}$ model and various associated expressions in terms of the $R_5$ scale radius and $\mu_5$, and the fraction of total light within $R_5$ is reported in Section~\ref{Sec_frac}. 
This reformulation allows several standard quantities --- the central
surface brightness, the mean surface brightness within $R_5$, the
enclosed-light curve of growth, and the non-parametric Petrosian and Kron
radii --- to be re-expressed as direct, closed-form functions of $n$,
bypassing the transcendental equation that defines $b_n$
(Section~\ref{Sec_Assoc}).

The deprojection analysis (Section~\ref{Sec_depro-nu} and Appendix~\ref{Sec_Appdx_A}) yields an exact single-integral relation between projected (2D) and spherically enclosed (three-dimensional: 3D) light fractions when sharing the same radius, avoiding approximate deprojections when converting between the two via nested double integrals (Section~\ref{Sec_singleR2}). 
The deprojection of the $R_5$ scale also provides a dynamical
connection.  For a spherical $R^{1/n}$ luminosity-density profile, the
internal radius $r_{-3}$, where the logarithmic slope of the deprojected, spatial (3D) luminosity density profile equals
$-3$, is shown to be related to $R_5\equiv R_{-2}$ by a weakly $n$-dependent function, and 
as $n\rightarrow \infty$, the ratio $r_{-3}/R_{-2} \rightarrow \rm {e}/2$ (Section~\ref{Sec-r3-exp-R2} and Appendix~\ref{Appdx_Sec_r3_asymptotic}). 
This mapping connects the observable projected scale radius $R_5$ to the radius $r_{-3}$ where the mass--anisotropy degeneracy in the spherical Jeans equation 
is minimized \citep{2010MNRAS.406.1220W}, enabling a refined, $n$-dependent Wolf-type estimator
for the enclosed dynamical mass (Section~\ref{Sec_massive}) and its corresponding spatial mass-to-light ratio (Section~\ref{Sec_mottle}).  
Some past dynamical mass estimators are compared in Section~\ref{Sec_prior_mass}, and  
mass estimates for stellar systems free of dark matter reviewed in Section~\ref{Sec_Bertin} before being extended in Section~\ref{Sec_GC97_ext}.

The $R_5$ radial scale (Section~\ref{Sec_illusion}) 
and associated dynamical estimators are examined across the observed range of S\'ersic indices, with particular 
attention to the low-$n$ systems characteristic of dwarf and UDGs, for which the conventional use of the projected radius 
$R_{\rm e}$ and the internal radius $r_{1/2}$ can introduce appreciable systematic differences. The implications for UDGs and other low-$n$ systems are discussed in Section~\ref{Sec_UDGs}, with a worked example provided in Section~\ref{Sec_DF2_example} for NGC~1052–DF2. 
In these systems, the slope of the deprojected luminosity-density at the $r_{1/2}$ radius adopted in the Wolf mass estimator can differ substantially from the anisotropy-insensitive value of $-3$. 
Section~\ref{Sec_Conclusions} provides a summary and conclusions. 

To facilitate immediate practical application and S\'ersic parameter catalog conversions without requiring readers to navigate the complete mathematical 
derivations, a standalone ``Observer's Cookbook'' summarizing key 
structural relations, deprojected sweet-spot radii, dynamical mass 
estimators, and spatial mass-to-light formulas is provided in 
Appendix~\ref{Appdx_cookbook} (Table~\ref{tab:cookbook}). In addition, 
all analytical conversions, numerical Abel deprojections, and dynamical 
estimators developed herein are implemented in the open-source Python 
package \textsc{R5Tools} \citep{Graham_R5Tools_2026}, freely available on 
GitHub\footnote{\url{https://github.com/A-Graham/R5Tools}} and 
permanently archived on Zenodo, alongside an interactive web interface 
at \url{https://r5tools.streamlit.app/}.

\begin{figure*}[ht]
\begin{center}
\includegraphics[angle=270, trim=10cm 9.0cm 2.2cm 8.cm, width=\columnwidth]{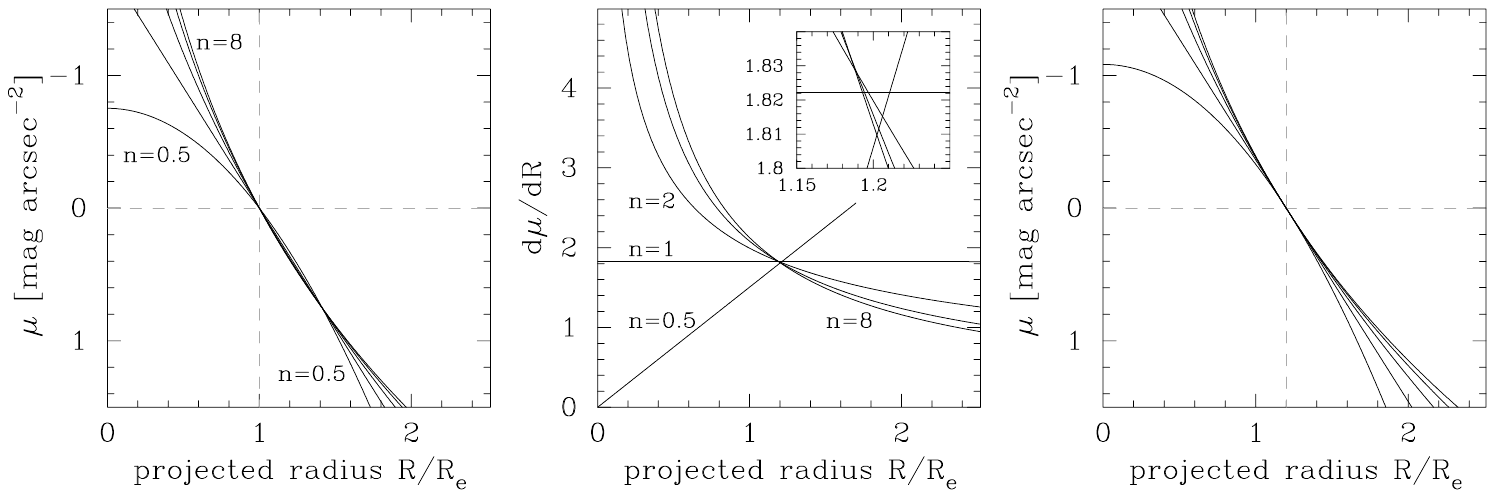}
\caption{Left panel: $R^{1/n}$ light profiles, with $n=0.5,1,2,4$ and 8, with 
  the surface brightness normalized to zero at $R=R_{\rm e}$. 
Middle panel: Linear gradient of the surface brightness profiles. 
As $n$ approaches infinity, the radius where the slope equals 1.822 approaches $1.181\,R_{\rm e}$ (Section~\ref{Sec_linear}). 
Right panel: $R^{1/n}$ light profiles normalized to a surface brightness of
zero at $R=1.19R_{\rm e}$, where the slope roughly equals 1.822 for different values of $n$.  
Shifts in the radius near $R\approx1.19R_{\rm e}$ are such that the associated shift in the surface brightness is given by $\Delta \mu \approx 1.822 (\Delta R)$ for $n\gtrsim0.5$. 
}
\label{Fig0A}
\end{center}
\end{figure*}

\section{A common slope and the \texorpdfstring{$R_5$}{Ah5} scale radius}

\subsection{The Linear Slope \texorpdfstring{$\mathrm{d}\mu/\mathrm{d}R$}{dmu/dR}}
\label{Sec_linear}

The surface brightness profile $\mu(R) \equiv -2.5\log\, I(R)$ that is associated with Equation~\ref{Eq_Ser} can be expressed as:
\begin{equation}
\mu(R) = \mu_{\rm e} + \frac{2.5 b_n}{\ln(10)} \left[ \left(\frac{R}{R_{\rm e}}\right)^{1/n} - 1 \right],
\end{equation}
with $\mu_{\rm e}$ known as the `effective surface brightness' at $R_{\rm e}$.  
Example surface brightness profiles are shown in the left hand panel of Figure~\ref{Fig0A},
where they have been normalized to a surface brightness of 0 mag~arcsec$^{-2}$ at $R=1\,R_{\rm e}$. 
It has long been known, and invariably ignored, that, intriguingly, things do not quite line up at $1\,R_{\rm e}$. 

The objective here is to find a radius near $R_{\rm e}$---which, below, is initially denoted $R_{\star}$---where these profiles do line up, where the slopes of the  profiles having different values of $n$ agree with each other; that is, to find a radius where the slope is largely independent of $n$. 

The linear slope of the surface brightness profile, $\mathrm{d}\mu/\mathrm{d}R$, is 
\begin{equation}
\frac{\mathrm{d}\mu(R)}{\mathrm{d}R} = \frac{2.5 b_n}{n \ln(10) R_{\rm e}} \left(\frac{R}{R_{\rm e}}\right)^{1/n - 1}.
\end{equation}
Evaluating this at a parameterized radius $R_{\star} = R_{\rm e} e^y$ yields:
\begin{equation}
\frac{\mathrm{d}\mu}{\mathrm{d}R} = \frac{2.5 b_n}{n \ln(10) R_{\rm e}} e^{-y} e^{y/n}.
\end{equation}

By definition, Capaccioli's binding parameter, $b_n$, is the exact solution to the median-defining equation $\gamma(2n, b_n) = \frac{1}{2}\Gamma(2n)$, involving the incomplete and complete gamma function \citep{1991A&A...249...99C}. This defines $b_n$ as the median of a Gamma distribution with a shape parameter of $2n$. In his early work, \citet{Ramanujan1911} \citep[see also][]{Ramanujan1912,Choi_1994,1999A&A...352..447C} posed a famous mathematical problem regarding the asymptotic expansion of this median, which relates the median $b_n$ to its mean $2n$ through the Laurent series:
\begin{equation}
b_n \approx 2n - \frac{1}{3} + \frac{4}{405n} + \mathcal{O}(n^{-2}),
\label{Eq_Ramanujan} 
\end{equation}
where the first-order approximation $b_n \approx 2n - 1/3 + 0.009876/n$ provides an
accurate description for $n \ge 0.5$ \citep[][their equation~A3a]{1997A&A...321..111P}.


Applying the asymptotic expansions ($b_n \approx 2n - 1/3$ and $e^{y/n} \approx 1 + y/n$) for large $n$:
\begin{equation}
\frac{\mathrm{d}\mu}{\mathrm{d}R} \approx \frac{2.5}{\ln(10) R_{\rm e}} e^{-y} \left(2 - \frac{1}{3n}\right)\left(1 + \frac{y}{n}\right) \approx \frac{2.5}{\ln(10) R_{\rm e}} e^{-y} \left[ 2 + \frac{2y - 1/3}{n} \right].
\end{equation}
Setting the first-order $1/n$ variation to zero, to remove the gradient's dependence on $n$, requires:
\begin{equation}
2y - \frac{1}{3} = 0 \implies y = \frac{1}{6} \implies \mathbf{R_{\star} = R_{\rm e} e^{1/6} \approx 1.181 R_{\rm e}}.
\label{Eq_one-sixth}
\end{equation}

This is approximately the radius, in units of half light radii, where $R^{1/n}$ light profiles have the same slope, independent of $n$.  
The slope at this radius is approximately 1.822 as $n\rightarrow \infty$. 

Another approach to this problem is to note that the slope equals $\approx$1.822 at all radii if $n=1$ (reproducing an exponential model), but varies with radius for other values of $n$.  Therefore, finding the radius where the slope is $\approx$1.822, independent of the value of $n$, should arrive at a similar answer close to $\approx1.181\,R_{\rm e}$. Indeed, Figure~\ref{Fig0A} reveals a value around 1.19$R_{\rm e}$.  Upon inspecting the surface brightness profile on the less commonly used logarithmic radial scale, the actual nexus point, and its significance, emerges more clearly.

\subsection{The logarithmic slope \texorpdfstring{$\mathrm{d}\mu/\mathrm{d}\log R$}{dmu/dR}}
\label{Sec_log}

\begin{figure*}[ht]
\begin{center}
\includegraphics[angle=270, trim=10cm 9.0cm 2.2cm 8.cm, width=1.0\columnwidth]{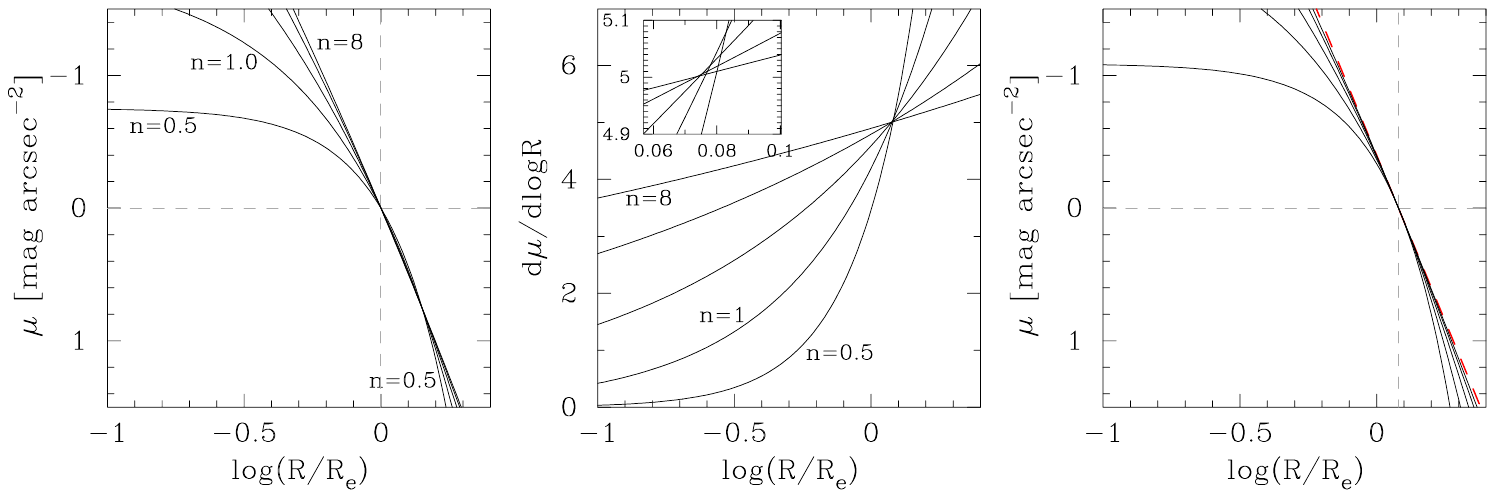}
\caption{Similar to Figure~\ref{Fig0A}, except that the radial coordinate is now
  on a logarithmic scale.  
At radii near 1.19$R_{\rm e}$, changes in $\log R$ are such that changes in the surface brightness $\Delta \mu \approx 5 (\Delta \log R)$ (or $\Delta(\log I) \approx -2 (\Delta \log R)$). 
The dashed red line in the right panel has a slope of 5. 
}
\label{Fig0B}
\end{center}
\end{figure*}

The same $R^{1/n}$ surface brightness profiles shown in Figure~\ref{Fig0A} are plotted on a logarithmic
radial scale in the left-hand panel of Figure~\ref{Fig0B}.  
Re-normalizing the surface brightness profiles to
$0\,\text{mag\,arcsec}^{-2}$ at $1.19R_{\rm e}$ in the right hand
panel of Figure~\ref{Fig0B}, one can see how the slopes of the
profiles with different S\'ersic indices are roughly equal (to a value
of 5) at $\log(R/R_{\rm e}) \approx 0.08$ ($R\approx1.19\, R_{\rm
  e}$).  If one over-estimates the value of $1.19 R_{\rm e}$, then the
error in $\mu_{\rm 1.19 R_{\rm e}}$ will roughly be the same for all
values $n$.  Similarly, if one under-estimates $1.19 R_{\rm e}$, then
the error in $\mu_{\rm 1.19 R_{\rm e}}$ will again roughly be the same
for all $n$.  More specifically, a measurement error in the value of
$\log(1.19R_{\rm e})$ (up to $\sim0.1$ dex, i.e.\ 25 per cent in
radius) will result in a measurement error in the surface brightness
that is roughly 5 times this value.  This is true for any light
profile that is well described by the $R^{1/n}$ model, and is now
tackled more precisely.

The local logarithmic slope, $S$, of the $R^{1/n}$ profile is given by the derivative:
\begin{equation}
S(R, n) \equiv \frac{\mathrm{d}\mu}{\mathrm{d}\log_{10} R} = \frac{2.5 b_n}{n}
\left(\frac{R}{R_{\rm e}}\right)^{1/n}, 
\label{Eq_log_slope}
\end{equation}
as per \citet[][their Eq.~23]{2005PASA...22..118G}. 
Now, the objective is to find an optimal logarithmic radius, $\log R_{\star\star}$, where the slope $S(R_{\star\star}, n)$ is independent of the S\'ersic index $n$. To find this, the radius is again parameterized as $R_{\star\star} = R_{\rm e} e^y$, where $y \equiv \ln(R_{\star\star}/R_{\rm e})$, which yields:
\begin{equation}
S(R_{\star\star}, n) = \frac{2.5 b_n}{n} {\rm e}^{y/n}.
\end{equation}

Using the above asymptotic expansion for the binding parameter $b_n$ for larger S\'ersic indices ($b_n \approx 2n - \frac{1}{3}$), one can write the slope as:
\begin{equation}
S(R_{\star\star}, n) \approx 2.5 \left(2 - \frac{1}{3n}\right) {\rm e}^{y/n}.
\end{equation}
Approximating the exponential term for a radius close to the effective radius ($e^{y/n} \approx 1 + \frac{y}{n} + \mathcal{O}(n^{-2})$) and keeping only the first-order terms, one obtains:
\begin{equation}
S(R_{\star\star}, n) \approx 2.5 \left(2 - \frac{1}{3n}\right)\left(1 + \frac{y}{n}\right) \approx 2.5 \left[ 2 + \frac{2y - 1/3}{n} \right].
\end{equation}
As before, to mathematically eliminate the first-order dependence on $n$, the coefficient of the $1/n$ term must vanish:
\begin{equation}
2y - \frac{1}{3} = 0 \implies y = \frac{1}{6}.
\end{equation}
%
%
Because the linear and logarithmic derivatives are connected by the chain rule ($\mathrm{d}\mu/\mathrm{d}R = \{1/[R \ln(10)]\}\,  \mathrm{d}\mu/\mathrm{d}\log R$), the logarithmic slope's first-order S\'ersic-index dependence is minimized at the exact same physical radius.
Evaluating the chain rule at $R = R_{\rm e}e^y$ 
introduces the factor $1/R = e^{-y}/R_{\rm e}$, which accounts for 
the $e^{-y}$ present in the linear slope but absent from 
$S(R_{\star\star}, n)$. 
However, what is now apparent is that when $2y-1/3 = 0$, the logarithmic slope $S(R_{\star\star}, n) \approx 5$. 

Having identified a convergence on the radius where the logarithmic slope is $\approx$5, the radius where the slope is exactly 5 is now introduced. 
Denoting this as $R_5$, Equation~\ref{Eq_log_slope} can be written as 
\begin{equation}
R_5 = \left( \frac{2n}{b_n} \right)^n R_{\rm e}. 
\label{Eq_R5}
\end{equation}  
It is believed that this is the first time an exact analytical expression is presented for the radius at which the local logarithmic slope of an $R^{1/n}$ surface brightness profile equals 5.
Equation~\ref{Eq_R5} reveals how the ratio $R_5/R_{\rm e}$ shifts smoothly from $\approx 1.2$ at $n=0.5$ down to the asymptotic limit of $e^{1/6} \approx 1.181$ as $n \to \infty$.  This is reported in Table~\ref{Tab_R5_ratios} for select values of $n$.

\begin{table}[h]
\centering
\caption{Columns~1 and 2: S\'ersic index and associated binding parameter $b_n$. 
Column~3: Ratio of the projected scale radius $R_{5}$ to the effective half-light radius $R_{\rm e}$. 
Column~4: Fraction of the total light that is projected within $R_5$, equal to 
$\gamma(2n, 2n)/\Gamma(2n)$. 
Columns~5 and 6: Non-parametric concentration index $\mathcal{F}_{0.33/5}(n)$ (Equation~\ref{eq:R_bn_free}) and, for convenience, the total luminosity 
extrapolation factor $\mathcal{F}_{\infty/5}(n)= \Gamma(2n)/\gamma(2n,\,2n)$ (Equation~\ref{Eq_bigF}). 
Column~7 displays $\Delta\mu \equiv \mu_5-\mu_{\rm e}$. 
The $n=\infty$ row gives the exact analytic limits, approached only asymptotically. 
}
\label{Tab_R5_ratios}
\begin{tabular}{rrccccc}
\hline
$n$ &  $b_n$  &   $R_{5}/R_{\rm e}$ & $L_{\rm 2D}(<R_5)$ &  $\mathcal{F}_{0.33/5}$ & $\mathcal{F}_{\infty/5}$ & $\Delta\mu$ \\
  (1) &   (2)             &        (3)      &    (4)  &     (5)          &   (6)     &  (7)      \\
\hline
0.25                   & 0.2275         & 1.2176  & $0.6827$  &  $0.130$ & $1.465$  & 0.296 \\
0.5                    & 0.6932         & 1.2011  & $0.6321$  &  $0.166$ & $1.582$  & 0.333 \\
1.0                    & 1.6784         & 1.1916  & $0.5940$  &  $0.243$ & $1.684$  & 0.349 \\
2.0                    & 3.6721         & 1.1866  & $0.5665$  &  $0.358$ & $1.765$  & 0.356 \\
3.0                    & 5.6702         & 1.1849  & $0.5543$  &  $0.434$ & $1.804$  & 0.358 \\
4.0                    & 7.6693         & 1.1840  & $0.5470$  &  $0.488$ & $1.828$  & 0.359 \\
5.0                    & 9.6687         & 1.1835  & $0.5421$  &  $0.529$ & $1.845$  & 0.360 \\
6.0                    & 11.6680        & 1.1831  & $0.5384$  &  $0.562$ & $1.857$  & 0.360 \\
7.0                    & 13.6681        & 1.1829  & $0.5355$  &  $0.588$ & $1.867$  & 0.360 \\
8.0                    & 15.6679        & 1.1827  & $0.5333$  &  $0.610$ & $1.875$  & 0.361 \\
9.0                    & 17.6678        & 1.1825  & $0.5313$  &  $0.629$ & $1.882$  & 0.361 \\
10.0                   & 19.6677        & 1.1824  & $0.5297$  &  $0.645$ & $1.888$  & 0.361 \\
$\rightarrow \infty$   & $\rightarrow \infty$ & 1.1814 & $0.5000$ & $1.000$ & $2.000$ & 0.362 \\
\hline
\end{tabular}
\end{table}

The surface brightness at $R_5$ is related to the effective-surface brightness at $R_{\rm e}$ by the equation 
\begin{equation}
\mu_5=\mu_{\rm e}-\frac{2.5}{\ln 10}\,(b_n-2n).
\label{Eq_mu5_mue}
\end{equation}
The difference \(\mu_5-\mu_{\rm e}\) is always positive and varies only 
weakly with S\'ersic index (Table~\ref{Tab_R5_ratios}), rising from 
0.296\,mag at \(n=0.25\) to the asymptotic value \(2.5/(3\ln 10)\approx0.362\)\,mag 
as \(n\to\infty\) (via Equation~\ref{Eq_Ramanujan}).  Consequently, once \(R_5\) and \(\mu_5\) have been measured, 
the conventional parameters \(R_{\rm e}\) and \(\mu_{\rm e}\) can readily be recovered.

The radius $R_5$ is equivalently the radius where $\mathrm{d}\ln I / \mathrm{d}\ln R \equiv \mathrm{d}\log I / \mathrm{d}\log R$ equals $-$2, and to help with clarity, the designation $R_{-2}$ is also used in some of what follows.

\section{Some benefits of using \texorpdfstring{$R_5$}{R5} and 
\texorpdfstring{$\mu_5$}{mu5}}
\label{Sec_benefit}

The radius $R_5$ offers several practical and conceptual advantages over 
$R_{\rm e}$ as a photometric reference scale.

First, because $m_{\rm tot} \propto \mu_5 - 5\log R_5$, 
perturbations in $R_5$ and $\mu_5$  analytically compensate to first order when 
computing the extrapolated total magnitude 
(Section~\ref{Sec_stable}).

Second, $R_5$ is the projected radius at which the luminosity contribution 
per logarithmic interval of radius, $\mathrm{d}L/\mathrm{d}\ln R \propto R^2 I(R)$, reaches 
its global maximum (Section~\ref{Sec_benefit_phys}). This gives $R_5$ a 
natural physical interpretation as the most luminosity-weighted scale in 
the projected light distribution.

Third, $R_5$ can be identified directly from the observed surface brightness 
profile as the point where the logarithmic slope equals  
$5.00\,\text{mag\,dex}^{-1}$, without needing to fit an $R^{1/n}$ model 
(Section~\ref{Sec_Non-par}). Once identified, it provides a 
fitting-free route to the total magnitude via the concentration index 
$\mathcal{F}_{0.33/5}$ and extrapolation factor $\mathcal{F}_{\infty/5}(n)$ (Table~\ref{Tab_R5_ratios}).
Furthermore, because $R_5$ can be measured as a local gradient in the high-signal-to-noise 
region of the profile, its measurement is less sensitive to sky-background 
subtraction errors and when the faint outer wings are truncated by surface-brightness 
limits.   When fit parametrically, it is less sensitive to the `seeing' than the smaller radius $R_{\rm e}$.

Fourth, parameterizing the $R^{1/n}$ profile in terms of $R_5$ eliminates 
the binding parameter $b_n$ and the half-light radius $R_{\rm e}$ from 
the analytical model entirely (Section~\ref{Sec_new_rel}), 
substantially reducing the mathematical coupling among the 
model parameters.

Finally, $R_5$ can be  connected to the dynamically robust radius 
$r_{-3}$, where the logarithmic slope of the deprojected luminosity density 
equals $-3$ and the mass-anisotropy degeneracy in the spherical Jeans 
equation is minimized. This connection enables a refined, $n$-dependent 
dynamical mass estimator expressed in terms of the observable $R_5$ 
(Section~\ref{Sec_massive}).

\subsection{Decoupling S\'ersic Parameters and Error Compensation at \texorpdfstring{$R_5$}{R5}}
\label{Sec_stable}

Parameterizing the $R^{1/n}$ profile in terms of $(R_5,I_5)$ 
addresses a notable analytical and numerical limitation
of the standard 
$(R_{\rm e},I_{\rm e})$ variables. In the standard formulation of the $R^{1/n}$ model, the effective 
half-light radius $R_{\rm e}$ is a non-local, integral quantity related to the  
integration of the light profile from the centre to infinity. The non-linear 
coupling term $b_n(n)$  tightly binds the spatial scale $R_{\rm e}$, 
the intensity scale $I_{\rm e}$ and the shape parameter $n$. During non-linear 
fitting, and multi-component decompositions, any uncertainty in $n$ forces a 
highly correlated adjustment in both $R_{\rm e}$ and $I_{\rm e}$, producing 
the well-known parameter degeneracies 
\citep[e.g.,][]{1996ApJ...465..534G,2001MNRAS.326..869T}.

Transitioning to a $b_n$-free form reduces parameter coupling in the
analytical model by replacing the integral-defined $R_{\rm e}$ with the
gradient-defined scale radius $R_5$. Because $R_5$ is anchored to a local
landmark—specifically, the unique radius where the projected logarithmic
intensity slope is exactly $-2$—its mathematical definition is structurally
independent of the global curvature parameterized by $n$. By definition, the
local logarithmic slope of the surface-brightness profile at $R_5$ is exactly
$5.00\,\mathrm{mag\,dex^{-1}}$ for all values of $n$ (Figure~2). Consequently,
observational or fitting error that shift $\log R_5$ along the one-dimensional
profile by $\Delta\log R_5$ induces a compensating shift
$\Delta\mu_5\approx5\,\Delta\log R_5$. 
This error vector is parallel to the
line of constant total magnitude, since $m_{\rm tot}\propto\mu_5-5\log_{10}R_5$,
leaving the inferred total luminosity $L_{\rm tot}$ invariant to first order.   
This desirable behaviour holds for any 
S\'ersic profile with $n \gtrsim 0.5$, and for radius errors of up to 
$\sim$30 per cent (Figure~\ref{Fig0B}).
Although, in practical two-dimensional galaxy fitting subject
to noise, point spread function (PSF) blurring and non-S\'ersic structural components, statistical
covariances among parameters cannot be fully eliminated by such a coordinate
transformation alone. 
Full end-to-end recovery tests with realistic noise, sky errors and multi-component profiles are, however, beyond the scope of the present analytical study and are deferred to future work.

\subsection{Some physical Significance of the Scale Radii \texorpdfstring{$R_{5}$}{R\_{5}}}
\label{Sec_benefit_phys}

To understand the physical significance of the projected scale radius $R_5$, the distribution of projected luminosity on the plane of the sky is examined. The projected luminosity contained within a thin circular annulus of radius $R$ and physical width $\mathrm{d}R$ is given by:
\begin{equation}
\mathrm{d}L = I(R) dA = 2\pi R I(R) \, \mathrm{d}R.
\end{equation}

By dividing the galaxy into logarithmic intervals of radius ($\mathrm{d}\ln R = \mathrm{d}R/R$), the projected luminosity contribution per logarithmic interval is:
\begin{equation}
\frac{\mathrm{d}L}{\mathrm{d}\ln R} = R \frac{\mathrm{d}L}{\mathrm{d}R} = 2\pi R^2 I(R).
\end{equation}
To find the radius $R_{\rm peak, log}$ where this logarithmic luminosity distribution is maximized, its derivative with respect to $R$ is set to zero: 
\begin{equation}
\frac{\mathrm{d}}{\mathrm{d}R}\left[ R^2 I(R) \right] = 2R I(R) + R^2 \frac{\mathrm{d}I}{\mathrm{d}R} = 0.
\end{equation}
Dividing by $R I(R)$ and rearranging terms yields:
\begin{equation}
\frac{R}{I(R)} \frac{\mathrm{d}I}{\mathrm{d}R} \equiv \frac{\mathrm{d}\ln I}{\mathrm{d}\ln R} = -2.
\end{equation}
This mathematically proves that the $R_5$ scale radius (a.k.a.\ $R_{-2}$, the radius where the $\mathrm{d}\ln I/\mathrm{d}\ln R = -2$) is the exact projected radius where the projected luminosity contribution per logarithmic interval of radius, $\mathrm{d}\ln R$, reaches its global maximum.\footnote{For constant $M/L$, the same radius also maximizes the projected stellar-mass contribution. The Author thanks Agris Kalnajs (priv.\ comm.\ 2004) for thinking this way.}

For comparison, if the distribution is instead analyzed in linear intervals of radius, the projected luminosity per unit physical radial interval ($\mathrm{d}R$) is:
\begin{equation}
\frac{\mathrm{d}L}{\mathrm{d}R} = 2\pi R I(R).
\end{equation}
To find the radius $R_{\rm peak, lin}$ where this physical shell distribution reaches its global maximum, the derivative with respect to $R$ is set to zero:
\begin{equation}
\frac{\mathrm{d}}{\mathrm{d}R}\left[ R I(R) \right] = I(R) + R \frac{\mathrm{d}I}{\mathrm{d}R} = 0 \implies \frac{\mathrm{d}\ln I}{\mathrm{d}\ln R} = -1.
\end{equation}
This shows that the luminosity in a projected circular shell of
constant physical width $\mathrm{d}R$ is maximized at the radius $R_{-1}$, where the
projected logarithmic intensity slope is exactly $-1$ (coinciding with the
classic scale-length $h$ when $n=1$). 
This is the \emph{mode} of the shell
distribution $R\,I(R)$, however, and is distinct from its \emph{mean}. For
an exponential profile ($n=1$), the centroid (first moment) of this same
shell distribution over the domain $[0,\infty)$ is located not at $R_{-1}=h$
but at exactly $2h$.\footnote{This identity is derived in Section~\ref{Sec_Assoc}.} As discussed in
Section~\ref{Sec_Assoc}, this centroid is exactly equal to the scale radius
$R_5$, which consequently matches the asymptotic Kron radius $R_1(\infty)
\equiv \int_0^\infty R^2 I(R)\,\mathrm{d}R \big/ \int_0^\infty R\,I(R)\,\mathrm{d}R$ in this
regime.

\subsection{A Non-Parametric Approach}
\label{Sec_Non-par}

\subsubsection{The Non-Parametric Scale Radius \texorpdfstring{$R_{5}$}{R5}}

An outstanding practical challenge in galaxy photometry is that the parameters from parametric 
S\'ersic fits---namely $R_{\rm e}$, $\mu_{\rm e}$, and $n$---are often 
highly degenerate when obtained via non-linear fitting algorithms, 
exacerbated by the point spread function (PSF) blurring, low signal-to-noise in the 
outer wings, or sky-subtraction errors.

The $R_5$ scale radius  can be measured without recourse to parametric fitting, 
by identifying it directly from the observed, azimuthally averaged 
surface-brightness profile, $\mu(R)$, as the point where the local logarithmic 
slope equals exactly $5.00\,\text{mag\,dex}^{-1}$:
\begin{equation}
\left. \frac{\mathrm{d}\mu}{\mathrm{d}\log_{10} R} \right|_{R=R_{5}} = 5.00.
\label{eq:Rstar_def}
\end{equation}
For profiles where the local logarithmic slope increases monotonically with 
radius (as is the case with the $R^{1/n}$ function), this point is unique. For 
galaxies well described by an $R^{1/n}$ model with $n \gtrsim 0.5$, this 
gradient-defined landmark corresponds to $R_{5} \approx 
1.18$--$1.20\,R_{\rm e}$ (Table~\ref{Tab_R5_ratios}), and thus $R_{\rm e} \approx (0.83$--$0.85)\,R_{5}$.

\subsubsection{Concentration and Extrapolated Luminosity}
\label{Sec_Non-par-2}

Once $R_5$ is identified from the observed profile via 
Equation~(\ref{eq:Rstar_def}), the global concentration of the galaxy can 
be quantified without fitting an $R^{1/n}$ model. A concentration index 
$\mathcal{F}_{0.33/5}$ is defined as the fraction of the projected luminosity enclosed 
within $(1/3)R_5$ to that within $R_5$:
\begin{equation}
\mathcal{F}_{0.33/5} \equiv \frac{L\left(<1/3\,R_5\right)}{L(<R_5)}.
\label{eq:R_ratio}
\end{equation}
Using the scale-radius relation $R_5 = (2n/b_n)^n R_{\rm e}$ 
(Equation~\ref{Eq_R5}), the standard integration variable,  
$x$, for the luminosity of the $R^{1/n}$ function within a given radius $R$ \citep[][their equation~2]{2005PASA...22..118G}, transforms as
\begin{equation}
x = b_n \left(\frac{R}{R_{\rm e}}\right)^{1/n} 
  = b_n \cdot \frac{(2n/b_n)(R/R_5)^{1/n}}{1} 
  = 2n\left(\frac{R}{R_5}\right)^{1/n},
\end{equation}
eliminating $b_n$ entirely from the enclosed-light expressions (Equation~\ref{eq:L_R_exact}).  
Consequently, the concentration ratio takes the $b_n$-free form:
\begin{equation}
\mathcal{F}_{0.33/5}(n) = \frac{\gamma\!\left(2n,\; 2n\cdot 3^{-1/n}\right)}
                      {\gamma\!\left(2n,\; 2n\right)},
\label{eq:R_bn_free}
\end{equation}
where $\gamma(a,x)$ is the lower incomplete gamma function. Since 
$\mathcal{F}_{0.33/5}(n)$ is a monotonically increasing function of $n$, the {\em observed}  
flux ratio can be used to uniquely infer the S\'ersic index without a fit, 
and serves as a stable analogue to the concentration index $C_{R_{\rm e}}(1/3)$ 
of \citet{2001MNRAS.326..869T}.\footnote{A concentration index formed with a more seeing-resistant inner aperture of $R_5/2$ is
\begin{equation}
\mathcal{F}_{0.5/5}(n)=\frac{L(<R_5/2)}{L(<R_5)}=\frac{\gamma\bigl(2n,\,2n\cdot(1/2)^{1/n}\bigr)}{\gamma(2n,\,2n)}.
\end{equation}
}

The total extrapolated luminosity then follows as
\begin{equation}
L_{\rm tot} = L(<R_5)\times\mathcal{F}_{\infty/5}(n),
\qquad
\mathcal{F}_{\infty/5}(n) = \frac{\Gamma(2n)}{\gamma(2n,\,2n)},
\label{Eq_bigF}
\end{equation}
where the flux factor $\mathcal{F}_{\infty/5}(n)$ can be read from a pre-computed look-up table indexed 
by $\mathcal{F}_{0.33/5}$. The procedure is therefore: measure $L(<R_5)$ and 
$L(< 1/3\, R_5)$ from the curve of growth, compute the fraction $\mathcal{F}_{0.33/5}$, look 
up the corresponding $n$ and $\mathcal{F}_{\infty/5}(n)$, and recover $L_{\rm tot}$---all 
without fitting an $R^{1/n}$ profile, although implicitly assuming that one provides a good description of the  
light distribution.

In practice, reliable non-parametric identification of $R_5$ requires that 
it lies well outside the PSF-affected region, i.e.\ $R_5 \gg r_{\rm PSF}$, 
where $r_{\rm PSF}$ is a characteristic PSF radius (e.g.\ half the FWHM). 
For nearby, well-resolved galaxies, this condition is easily satisfied. For 
compact or distant systems where $R_5$ approaches the PSF scale, the 
observed slope profile will be smoothed and the gradient-defined $R_5$ 
should instead be recovered from a PSF-deconvolved or model-assisted profile, 
at which point the method reduces to a constrained parametric fit, albeit measuring $R_5$, which is larger than $R_{\rm e}$.

\section{Eliminating \texorpdfstring{\NoCaseChange{\Reff}}{Re} and \texorpdfstring{\NoCaseChange{\bn}}{bn} from the \texorpdfstring{\NoCaseChange{\Rn}}{Rn} model}
\label{Sec_new_rel}

By utilizing the projected $R_5$ scale radius, one can eliminate both $R_{\rm e}$ and $b_n$ from the S\'ersic equation. From the definition of the $R_5$ scale radius:
\begin{equation}
R_{5} = \left(\frac{2n}{b_n}\right)^n R_{\rm e} \implies R_{\rm e} = R_{5} \left(\frac{b_n}{2n}\right)^n.
\end{equation}
Substituting this expression for $R_{\rm e}$ into the exponential term of Equation~\ref{Eq_Ser} yields:
\begin{equation}
\begin{aligned}
-b_n \left[ \left(\frac{R}{R_{\rm e}}\right)^{1/n} - 1 \right] 
&= -b_n \left[ \left(\frac{R}{R_{5} (b_n/2n)^n}\right)^{1/n} - 1 \right] \\
&= -2n \left(\frac{R}{R_{5}}\right)^{1/n} + b_n.
\end{aligned}
\end{equation}
Thus, the new form of the intensity profile can be rewritten as:
\begin{equation}
I(R) = I_{\rm e} e^{b_n} \exp \left[ -2n \left(\frac{R}{R_{5}}\right)^{1/n} \right] 
= I_0 \exp \left[ -2n \left(\frac{R}{R_{5}}\right)^{1/n} \right].
\label{eq:Sersic_new_I0}
\end{equation}

If the $R^{1/n}$ profile is defined in terms of the intensity $I_{5}$, such that 
\begin{equation}
I_{5} \equiv I(R_{5}) = I_0 e^{-2n}, 
\end{equation}
then one has
\begin{equation}
I(R) = I_{5} \exp \left\{ -2n \left[ \left(\frac{R}{R_{5}}\right)^{1/n} - 1 \right] \right\}, 
\label{eq:Sersic_new_I2}
\end{equation}
where $\mu_5 \equiv -2.5\log I_5$, and 
\begin{equation}
\frac{\mathrm{d}\ln I}{\mathrm{d} \ln R} = -2 \left(\frac{R}{R_5} \right) ^{1/n}.
\label{Eq_new_slope}
\end{equation}

Equations~(\ref{eq:Sersic_new_I0}) and (\ref{eq:Sersic_new_I2}) represent a significant conceptual simplification of the $R^{1/n}$ light profile. They depend only on the physical scale radius $R_{5}$, the S\'ersic index $n$, and the normalization intensity ($I_0$ or $I_{5}$), thereby completely bypassing the need to solve for or invoke the half-light-dependent parameters $R_{\rm e}$ and $b_n$.  Importantly, this form of the $R^{1/n}$ light profile model differs from \citet{1963BAAA....6...41S} due to the simple, but crucial, additional factor of 2 in the exponent.

\begin{figure}
\begin{center}
\includegraphics[angle=0, width=1.0\columnwidth]{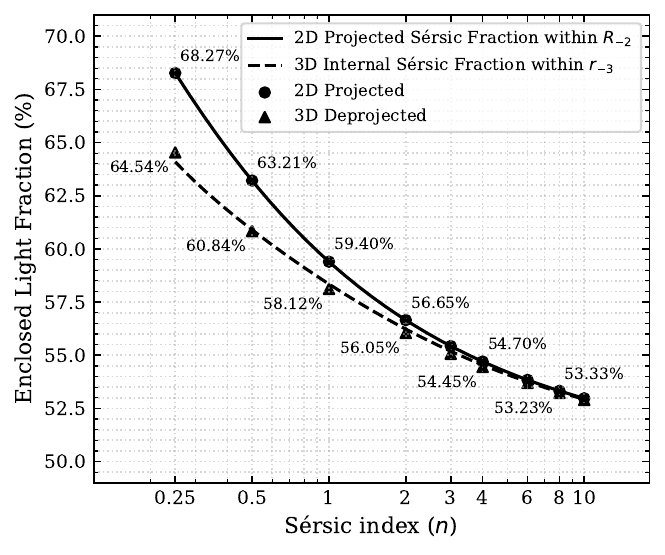}
\caption{Fraction of projected light contained within a cylindrical aperture of radius $R=R_5 \equiv R_{-2}$ and within a sphere of radius $r=r_{-3}$ for a range of $n$. 
The solid and dashed curves represent the analytical \citet{1963BAAA....6...41S} and 
\citet{1997A&A...321..111P} models, respectively, while the 
discrete circles (2D projected) and triangles (3D deprojected) indicate 
the exact, numerically integrated physical values (not from the Prugniel--Simien approximation).
}
\label{Fig_Sersic_fractions}
\end{center}
\end{figure}

\subsection{Associated expressions}
\label{Sec_Assoc}

The above $b_n$-free parameterization reveals a connection to the 
classic scale-length, $h$, that appears in the alternative expression  
$I(R)=I_0\exp[-(R/h)^{1/n}]$ \citep[e.g.,][their Eq.~14]{1988MNRAS.232..239D, 2005PASA...22..118G}. 
Equating the exponents of this expression with Equation~(\ref{eq:Sersic_new_I0}) immediately gives
\begin{equation}
R_5=(2n)^n\,h.
\label{eq:R5_scale_length}
\end{equation}
Equation~(\ref{eq:R5_scale_length}) completely bypasses both $b_n$ and the 
effective half-light radius. For a pure exponential ($n=1$), the relation neatly reduces to 
\begin{equation}
R_5=2h,
\end{equation}
showing that the $R_5$ radius of an exponential profile is precisely 
twice its traditional scale-length.

The surface brightness profile of the reformulated $R^{1/n}$ family, when 
reformulated in terms of $R_5$ and $I_5$, can be expressed as:
\begin{equation}
\mu(R) = \mu_0 + \frac{5n}{\ln(10)} \left(\frac{R}{R_{5}}\right)^{1/n} \approx \mu_0 + 2.1715 n \left(\frac{R}{R_{5}}\right)^{1/n},
\label{eq:mu_R_mu0}
\end{equation}
and
\begin{equation}
\mu(R) = \mu_{5} + \frac{5n}{\ln(10)} \left[ \left(\frac{R}{R_{5}}\right)^{1/n} - 1 \right] \approx \mu_{5} + 2.1715 n \left[ \left(\frac{R}{R_{5}}\right)^{1/n} - 1 \right],
\label{eq:mu_R_mu5}
\end{equation}
where $\mu_0$ is the central surface brightness and $\mu_{5}$ is the local surface brightness at $R_{5}$.

Evaluating Equation~\ref{eq:mu_R_mu5} at $R = 0$ yields a remarkably clean, 
$b_n$-free linear relationship for the central surface brightness:
\begin{equation}
\mu_0 = \mu_{5} - \frac{5n}{\ln(10)} \approx \mu_{5} - 2.1715\,n,
\label{eq:mu0_mu5}
\end{equation}
which serves as the direct, simplified counterpart to the classic relation 
$\mu_0 = \mu_{\rm e} - 2.5b_n/\ln(10)$ \citep[their Eq.~7]{2005PASA...22..118G}.

Similarly, the mean (average) surface brightness within the 
projected scale radius $R_5$, denoted $\langle \mu \rangle_5 \equiv -2.5\log_{10} 
\langle I \rangle_5$, can be computed. The average projected intensity, $\langle I \rangle_5$, 
within the circular aperture $A = \pi R_5^2$ is defined by integrating the intensity 
profile (Equation~\ref{eq:Sersic_new_I2}) to $R = R_5$:
\begin{equation}
\langle I \rangle_5 \equiv \frac{L(<R_5)}{\pi R_5^2} = I_5\, f_5(n),
\end{equation}
where the dimensionless intensity ratio $f_5(n)$ is defined as:
\begin{equation}
f_5(n) \equiv \frac{2n\, {\rm e}^{2n}\, \gamma(2n,\, 2n)}{(2n)^{2n}}.
\label{eq:f5_definition}
\end{equation}
Expressed in magnitude units, the mean surface brightness inside $R_5$ is 
given by the $b_n$-free expression:
\begin{equation}
\langle \mu \rangle_5 = \mu_{5} - 2.5\log_{10} \left[ f_5(n) \right],
\label{eq:mean_mu5}
\end{equation}
which replaces the traditional expression for the average surface brightness within
the effective radius \citep[their Eq.~9]{2005PASA...22..118G} --- itself requiring the
prior numerical solution of the transcendental equation defining $b_n(n)$ --- with a
single, direct evaluation of the incomplete gamma function.

The cumulative projected luminosity $L(<R)$ enclosed within a circular 
aperture of radius $R$ is obtained by integrating the intensity profile:
\begin{equation}
L(<R) = 2\pi \int_0^R I(R') R' \, \mathrm{d}R'.
\end{equation}
With $I(R) = I_0 \exp[-2n (R/R_5)^{1/n}]$ (Equation~\ref{eq:Sersic_new_I0}) and using the substitution 
$x' = 2n (R'/R_5)^{1/n}$
as seen in Section~\ref{Sec_Non-par-2}, 
and thus ${\rm d}x^{\prime}/{\rm d}R^{\prime} = 
x^{\prime}/(nR^{\prime})$, the integral becomes
\begin{equation}
L(<R) = \frac{2\pi n\, I_0}{(2n)^{2n}} R_{5}^2 \int_{0}^{2n(R/R_{5})^{1/n}} 
{\rm e}^{-x^{\prime}}(x^{\prime})^{2n-1} {\rm d}x^{\prime},
\end{equation}
which yields the exact expression
\begin{equation}
L(<R) = \frac{2\pi n\, I_0 R_5^2\, \gamma(2n,\, x)}{(2n)^{2n}} = \frac{2\pi n\, I_5 R_5^2\, e^{2n}\, \gamma(2n,\, x)}{(2n)^{2n}}, 
\label{eq:L_R_exact}
\end{equation}
where $x = 2n (R/R_5)^{1/n}$ and $\gamma(2n,x)$ is the lower incomplete 
gamma function.

The total (extrapolated) luminosity is recovered as $R\to\infty$ ($x\to\infty$, $\gamma(2n,x)\to\Gamma(2n)$):
\begin{equation}
L_{\rm tot} = \pi I_0 R_5^2 \frac{\Gamma(2n)}{(2n)^{2n-1}} = \pi I_5 R_5^2 \frac{e^{2n} \Gamma(2n)}{(2n)^{2n-1}}.
\label{eq:L_tot_exact}
\end{equation}

An interesting property of the complete Gamma function is that $\Gamma(2n+1)=(2n)! = (2n)\times(2n-1)\times (2n-2) \times \dots \times3\times2\times1$, for positive integer values of $n$, while $2n\,\Gamma(2n)=\Gamma(2n+1)$ for all positive real numbers.  As such, $\Gamma(2n)/(2n)^{2n-1} = \Gamma(2n+1)/(2n)^{2n}$. 

In magnitude units, the aperture magnitude (curve of growth) is
\begin{equation}
m(<R) = \mu_5 - 5\log_{10} R_5 - 2.5\log_{10} \left[ \frac{\pi e^{2n} \gamma(2n,\, x)}{(2n)^{2n-1}} \right],
\label{eq:m_R_mu5}
\end{equation}
and the total magnitude can be written as 
\begin{equation}
m_{\rm tot} = \mu_5 - 5\log_{10} R_5 - 2.5\log_{10} \left[ \frac{\pi e^{2n} \Gamma(2n+1)}{(2n)^{2n}} \right].
\label{eq:m_tot_mu5}
\end{equation}

The mean surface brightness within radius $R$ (in arcseconds) follows as
\begin{equation}
\langle \mu \rangle_R = m(<R) + 2.5 \log_{10} (\pi R^2).
\end{equation}

Finally, the classic non-parametric observational estimators can be 
translated into the $R_5$ coordinate system.  
The Petrosian function $\eta(R)$, defined as the ratio of the average 
surface brightness within projected radius $R$ to the local surface 
brightness at that radius \citep[their Eq.~30]{1976ApJ...209L...1P, 2005PASA...22..118G}, 
collapses to the $b_n$-free expression
\begin{equation}
\eta(R,\,n)=\frac{2n\,\gamma(2n,\,x)}{{\rm e}^{-x}\,x^{2n}},
\label{eq:R5_petrosian}
\end{equation}
where the dimensionless variable is again simply $x=2n(R/R_5)^{1/n}$. 
Because Equation~(\ref{eq:R5_petrosian}) is independent of $b_n$, the 
Petrosian radius $R_{\rm P}$ (defined by $\eta(R_{\rm P})=\eta_0$, where $\eta_0$ is a chosen threshold for the ratio of the average 
surface brightness within $R$ to the local surface brightness at $R$) is 
obtained as a clean ratio $R_{\rm P}/R_5$ that depends only on the 
S\'ersic index $n$.

Similarly, the luminosity-weighted first moment (Kron radius) 
\citep{1980ApJS...43..305K} takes the exact finite-aperture form
\begin{equation}
R_1(R,\,n)=\frac{\gamma\bigl(3n,\,2n(R/R_5)^{1/n}\bigr)}{(2n)^n\,\gamma\bigl(2n,\,2n(R/R_5)^{1/n}\bigr)}\,R_5
\label{eq:R5_kron_profile}
\end{equation}
\citep[cf.\ their Eq.~32]{2005PASA...22..118G}.  
In the limit $R\to\infty$ the incomplete gamma functions become complete 
and the asymptotic Kron radius reduces to the compact expression
\begin{equation}
R_1(\infty)=\frac{\Gamma(3n)}{(2n)^n\,\Gamma(2n)}\,R_5.
\label{eq:R5_kron_asymptotic}
\end{equation}

For a pure exponential profile ($n=1$) this yields the identity
\begin{equation}
R_1(\infty)=R_5,
\label{eq:R1_R5_identity}
\end{equation}
showing that the asymptotic Kron radius of an exponential profile is
exactly equal to $R_5$. This follows directly from the definition of the
Kron radius as the first moment (centroid) of the projected radial light
distribution, $R_1(\infty) \equiv \int_0^\infty R^2 I(R)\,\mathrm{d}R \big/
\int_0^\infty R\,I(R)\,\mathrm{d}R$. For an exponential profile this centroid lies
at exactly $2h$, identical to the scale radius $R_5$
(Equation~\ref{eq:R5_scale_length}).

\subsection{Enclosed Light Fractions within \texorpdfstring{$R=R_{5}$}{R\_{5}}}
\label{Sec_frac}

From Equations~\ref{eq:L_R_exact} and \ref{eq:L_tot_exact}, 
the exact fraction of total light enclosed within projected radius $R$ is
\begin{equation}
\frac{L(<R)}{L_{\rm tot}} = \frac{\gamma(2n,\, x)}{\Gamma(2n)}.
\label{eq:enclosed_fraction}
\end{equation}
At $R = R_5$, $x=2n$ and the above fraction is $\gamma(2n,2n)/\Gamma(2n)$, the inverse of Equation~\ref{Eq_bigF}, and $\approx 50$--$60$ per cent across the ETG sequence. 
The percentage of light contained within $R=R_5$ is shown in Figure~\ref{Fig_Sersic_fractions} and reported as a fraction in Table~\ref{Tab_R5_ratios} for different values of $n$.

As $n\to\infty$, the ratio $R_{5}/R_{\rm e}\to e^{1/6}\approx1.181$, and the 
projected light fraction within $R_{5}$ asymptotes to 50 per cent.\footnote{The 
asymptotic expansion of the regularized incomplete gamma function 
$P(a,a) \equiv \gamma(a,a)/\Gamma(a)$ for large $a$ is 
$P(a,a) \approx 1/2 + 1/(3\sqrt{2\pi a})$ \citep[e.g.,][section 11.2]{Temme1996}. 
For $a = 2n$, this yields the exact projected light fraction ratio 
$L(<R_5)/L_{\rm tot} \approx 1/2 + 1/(6\sqrt{\pi n})$, which converges to 
$1/2$ as $n \to \infty$.}
This occurs because, for very large $n$, the light profile becomes extremely centrally concentrated. Although $R_{5}$ lies $\sim$18 per cent beyond $R_{\rm e}$, the amount of light in the annular region between these two radii becomes negligible in the asymptotic limit.

\subsection{Prior scale radii associated with a slope of \texorpdfstring{$-2$}{-2}}

Assuming a constant mass-to-light ratio with radius, 
\citet{1965TrAlm...5...87E} deprojected the \citet{1948AnAp...11..247D}
$R^{1/4}$ model---building on the 
numerical deprojection tables of \citet{1960BOTT....2t...3P}---and found that
the resulting internal (3D) density profile was described, to good accuracy,
by the same general model but with a smaller exponent, $\nu \equiv 1/n \approx 0.18$ rather than 0.25.
\citet{1965TrAlm...5...87E} also noted that a generalized form of the $R^{1/4}$ model---that he proposed for describing the internal (3D) density of Galactic subsystems ---, in which the exponent $1/4$ is replaced by the parameter $\nu$, encompassed the special cases $\nu=1$ and $\nu=2$ 
(the exponential and Gaussian laws, respectively) that had already been used by other authors
for modelling internal density profiles.   
The expression in \citet[][his equation~4]{1965TrAlm...5...87E} used the
Sun's galactocentric radius as the scale radius rather than $r_{-2}$,
the internal radius where the logarithmic slope equals -2.\footnote{The coefficient $m_0$ in Einasto's exponential term was left as a free-fitted parameter (the log-slope at the solar radius) rather than fixed at $-2$, so the radius at
which the slope equals $-2$ does not in general coincide with the scale
radius of Einasto's original model.}

In other studies of internal 3D density profiles, the scale radius where the
logarithmic slope is exactly $-2$ was implicitly introduced as the
characteristic scale radius $r_{\rm s}$ of the double-power-law model of
\citet{1996ApJ...462..563N}. This concept was later applied by
\citet{2004MNRAS.349.1039N} to the three-parameter model originally
proposed by \citet{1965TrAlm...5...87E} to describe the internal mass
profiles of galaxies. \citet{2004MNRAS.349.1039N} explicitly defined the
internal scale-radius $r_{-2}$ as the location where $\mathrm{d}\ln\rho/\mathrm{d}\ln r
= -2$. Their equation~5 has the same functional form as used here, both
differing from the equation in \citet{1963BAAA....6...41S} due to the
crucial additional factor of 2 in the exponent (cf.\
Equations~\ref{eq:sersic_original} and \ref{eq:Sersic_new_I0}). Like the
internal profile of \citet{1965TrAlm...5...87E}, the equation in
\citet{2004MNRAS.349.1039N} was cast directly in terms of the 3D density;
unlike Einasto's own derivation, however, it was not obtained by
deprojecting a surface-density law, but was instead fit directly to
simulated halos. This parameterization of the internal (3D) density
profile was shown \citep{2004MNRAS.349.1039N, 2005ApJ...624L..85M,
2006AJ....132.2685M} to describe simulated dark matter halos better than
the NFW model \citep{1997ApJ...490..493N}, and was analyzed in detail by
\citet{2006AJ....132.2701G}.
That generalized equation, expressed in terms of the internal 3D scale 
radius $r_{-2}$, was, understandably, neither deprojected nor treated as a 2D distribution. 
As such, the expressions developed here are new.

\section{The Deprojection Connection: From \texorpdfstring{$R_{5}$}{R\_{5}} (a.k.a.\ \texorpdfstring{$R_{-2}$}{R\_{-2}}) to \texorpdfstring{\NoCaseChange{\rthree}}{r3}}
\label{sec_deproj}

\begin{table}
\centering
\caption{Column~1: S\'ersic index. 
Column~2: Fraction of the total light contained within the sphere of
internal radius $r=R_{-2}$, where the numerical value of the internal
radius is set equal to the projected scale radius $R_{-2}\equiv R_5$. 
Column~3: Difference between the fraction of the total light projected
within the circle of radius $R=R_{-2}$ (Column~4 of
Table~\ref{Tab_R5_ratios}) and that contained within the sphere of
internal radius $r=R_{-2}$ (Column~2), see Equation~\ref{eq:F3D_single_integral}.  
Column~4: Fraction of the total light contained within a sphere of radius $r = r_{-3}$. 
Column~5: Fraction of the total light that is projected within a cylinder/aperture of radius  $R=r_{-3}$. 
}
\label{tab:light_fractions}
\begin{tabular}{lcrcc}
\hline
$n$  & $L_{\rm 3D}$  &  $\Delta_{\rm 2D-3D}|_{R_{-2}}$   &  $L_{\rm 3D}$ & $L_{\rm 2D}$  \\
     & ($<r=R_{-2}$)    &                               &   ($<r=r_{-3}$)  &  ($<R=r_{-3}$) \\
(1)  &      (2)          &        (3)              &     (4)        &  (5)   \\
\hline
0.25             & 0.4304 &  0.2523 &  0.6454   & 0.8346 \\
0.5              & 0.4276 &  0.2045 &  0.6084   & 0.7769 \\
1.0              & 0.4370 &  0.1570 &  0.5812   & 0.7201 \\
2.0              & 0.4503 &  0.1162 &  0.5605   & 0.6688 \\
3.0              & 0.4578 &  0.0965 &  0.5507   & 0.6424 \\
4.0              & 0.4627 &  0.0843 &  0.5445   & 0.6256 \\
5.0              & 0.4663 &  0.0758 &  0.5402   & 0.6136 \\
6.0              & 0.4690 &  0.0694 &  0.5370   & 0.6045 \\
7.0              & 0.4711 &  0.0644 &  0.5344   & 0.5973 \\
8.0              & 0.4729 &  0.0604 &  0.5323   & 0.5914 \\
9.0              & 0.4743 &  0.0570 &  0.5305   & 0.5865 \\
10.0             & 0.4756 &  0.0541 &  0.5290   & 0.5823 \\
$\rightarrow \infty$ & 0.5000 & 0.0000 & 0.5000 & 0.5000 \\
\hline
\end{tabular}
\end{table}

\begin{table*}
\centering
\caption{Column~1: S\'ersic index.
Columns~2, 3, and 4: Logarithmic slope $\gamma_*$ of the deprojected stellar luminosity density evaluated at 
the internal radii whose numerical
values equal the projected scales $R_{\rm e}$ and $R_{-2}$, and at the
internal half-light radius $r_{1/2}$; that is, at
$r=R_{\rm e}$, $r=R_{-2}\approx1.19R_{\rm e}$, and
$r=r_{1/2}\approx1.33R_{\rm e}$, respectively. 
Columns~5, 6, and 7: Internal 3D radius $r_{-3}$, where the
deprojected stellar luminosity density has logarithmic slope $-3$,
expressed as the dimensionless ratios $r_{-3}/R_{\rm e}$,
$r_{-3}/R_{-2}$, and $r_{-3}/r_{1/2}$, respectively.
Columns~5 and 6 are also referred to as
$f(n)\equiv r_{-3}/R_{\rm e}$ and
$g(n)\equiv r_{-3}/R_{-2}$.  
For convenience, $r_{1/2}/R_{\rm e}$ and $r_{1/2}/R_{-2}$ are provided in Columns~8 and 9.
}
\label{tab:sersic_slopes}
\begin{tabular}{lcccccccc}
\hline
$n$  & $\gamma_*(r=R_{\rm e})$ & $\gamma_*(r = R_{-2})$  &  $\gamma_*(r=r_{1/2})$ & $r_{-3}/R_{\rm e}$ & $r_{-3}/R_{-2}$ & $r_{-3}/r_{1/2}$
  &  $r_{1/2}/R_{\rm e}$ & $r_{1/2}/R_{-2}$ \\
(1) & (2) & (3) & (4) & (5) & (6) & (7) & (8) & (9)\\
\hline
$0.25$ & $-0.24$ & $-1.22$ & $-1.70$ & $1.434$ & $1.178$ & 1.124 & $1.276$ & $1.048$ \\
$0.5$  & $-1.39$ & $-2.00$ & $-2.37$ & $1.471$ & $1.225$ & 1.126 & $1.306$ & $1.088$ \\
$1.0$  & $-2.13$ & $-2.46$ & $-2.68$ & $1.511$ & $1.268$ & 1.137 & $1.329$ & $1.115$ \\
$2.0$  & $-2.55$ & $-2.71$ & $-2.84$ & $1.547$ & $1.303$ & 1.153 & $1.342$ & $1.130$ \\
$3.0$  & $-2.69$ & $-2.80$ & $-2.89$ & $1.563$ & $1.319$ & 1.161 & $1.346$ & $1.136$ \\
$4.0$  & $-2.77$ & $-2.85$ & $-2.92$ & $1.572$ & $1.327$ & 1.165 & $1.349$ & $1.139$ \\
$5.0$  & $-2.81$ & $-2.88$ & $-2.93$ & $1.578$ & $1.333$ & 1.169 & $1.350$ & $1.141$ \\  
$6.0$  & $-2.84$ & $-2.90$ & $-2.95$ & $1.582$ & $1.337$ & 1.171 & $1.351$ & $1.142$ \\
$7.0$  & $-2.87$ & $-2.91$ & $-2.95$ & $1.585$ & $1.340$ & 1.172 & $1.352$ & $1.143$ \\
$8.0$  & $-2.88$ & $-2.92$ & $-2.96$ & $1.586$ & $1.342$ & 1.173 & $1.352$ & $1.144$ \\ 
$9.0$  & $-2.90$ & $-2.93$ & $-2.96$ & $1.589$ & $1.344$ & 1.175 & $1.352$ & $1.144$ \\
$10.0$ & $-2.91$ & $-2.94$ & $-2.97$ & $1.591$ & $1.345$ & 1.176 & $1.353$ & $1.144$ \\
$\rightarrow \infty$ & $-3.00$ & $-3.00$  & $-3.00$ & $1.607$ & $1.359$ & 1.185 & $1.356$ & $1.148$ \\ 
\hline
\end{tabular}
\end{table*}

\subsection{The deprojected (internal) luminosity density profile}
\label{Sec_depro-nu}

The 3D internal structure, that is, the spatial luminosity density profile, $\nu(r)$, of the $R^{1/n}$ model is obtained by solving the Abel integral equation \citep{1991A&A...249...99C}:
\begin{equation}
\nu(r) = -\frac{1}{\pi} \int_r^\infty \frac{\mathrm{d}I(R)/\mathrm{d}R}{\sqrt{R^2 - r^2}} \, \mathrm{d}R.
\label{eq:exact_deproj}
\end{equation}  
Taking the derivative of the intensity profile with respect to $R$ yields:
\begin{equation}
\frac{\mathrm{d}I(R)}{\mathrm{d}R} = I_5 e^{2n} \exp \left[ -2n \left(\frac{R}{R_5}\right)^{1/n} \right] \times \left[ -2n \cdot \frac{1}{n} \left(\frac{R}{R_5}\right)^{1/n - 1} \cdot \frac{1}{R_5} \right].
\end{equation}
Simplifying the terms inside the brackets, and swapping their order:
\begin{equation}
\frac{\mathrm{d}I(R)}{\mathrm{d}R} = -\frac{2 I_5 e^{2n}}{R_5} \left(\frac{R}{R_5}\right)^{1/n - 1} \exp \left[ -2n \left(\frac{R}{R_5}\right)^{1/n} \right],
\end{equation}
which can be grouped in terms of $R$ as:
\begin{equation}
\frac{\mathrm{d}I(R)}{\mathrm{d}R} = -\frac{2 I_5 e^{2n}}{R_5^{1/n}} R^{1/n - 1} \exp \left[ -2n \left(\frac{R}{R_5}\right)^{1/n} \right].
\label{eq:dIdR_R5}
\end{equation}
Inserting this derivative (Equation~\ref{eq:dIdR_R5}) into the Abel integral equation (Equation~\ref{eq:exact_deproj}) yields:
\begin{equation}
\nu(r) = -\frac{1}{\pi} \int_r^\infty \frac{(-2 I_5 e^{2n}/R_5^{1/n}) R^{1/n - 1} \exp \left[ -2n \left(R/R_5\right)^{1/n} \right]}{\sqrt{R^2 - r^2}} \, \mathrm{d}R,
\end{equation}
which simplifies directly to the exact integral expression:
\begin{equation}
\nu(r) = \frac{2 I_5 e^{2n}}{\pi R_5^{1/n}} \int_r^\infty \frac{R^{1/n - 1} \exp \left[ -2n \left(R/R_5\right)^{1/n} \right]}{\sqrt{R^2 - r^2}} \, \mathrm{d}R.
\label{eq:exact_deproj_R5_unsub}
\end{equation}
By defining a dimensionless radial integration variable, $y \equiv R/R_5$ (such that $\mathrm{d}R = R_5\mathrm{d}y$), 
Equation~(\ref{eq:exact_deproj_R5_unsub}) further simplifies to a clean, dimensionless integral that is computationally stable:
\begin{equation}
\nu(r) = \frac{2 I_5 e^{2n}}{\pi R_5} \int_{r/R_5}^\infty \frac{y^{1/n - 1} \exp \left( -2n y^{1/n} \right)}{\sqrt{y^2 - (r/R_5)^2}} \, \mathrm{d}y.
\label{eq:exact_deproj_R5_dimensionless}
\end{equation}

\subsection{Enclosed light fractions within \texorpdfstring{$r=R_{-2}$ and $r=r_{-3}$}{r = R-2 and r = r-3}}
\label{Sec_single}

\subsubsection{The projected radius $R_{-2}$}
\label{Sec_singleR2}

Integrating the internal density profile spherically yields the cumulative spatial (3D) luminosity profile $L(r) = 4\pi \int_0^r \nu(r') r'^2 \mathrm{d}r'$. 
While substituting the Abel inversion (Equation~\ref{eq:exact_deproj_R5_dimensionless}) directly 
into this expression produces a nested double integral, exchanging the order of 
integration simplifies the system. Because $I(R)$ is positive, smooth, and 
decays exponentially at large radii, absolute convergence is guaranteed and 
the order of integration may therefore be exchanged by Fubini's theorem. 

Reversing the order of integration and performing integration by parts on the 
radial derivative $\mathrm{d}I/\mathrm{d}R$ allows the inner integrals to be evaluated 
analytically (Appendix~\ref{Sec_Appdx_A}). The nested double integral collapses to a single-integral 
identity, defining the internal 3D light fraction $F_{\rm 3D}(r) \equiv 
L(r)/L_{\rm total}$ as:
\begin{equation}
F_{\rm 3D}(r) = F_{\rm 2D}(R=r) + \frac{4}{L_{\rm total}} \int_r^\infty I(R) R \left[ \arcsin\left(\frac{r}{R}\right) - \frac{r}{\sqrt{R^2 - r^2}} \right] \, \mathrm{d}R,
\label{eq:F3D_single_integral}
\end{equation}
where $F_{\rm 2D}(r) \equiv L_{\rm 2D}(R<r)/L_{\rm total}$ is the standard 
projected 2D light fraction (Equation~\ref{eq:enclosed_fraction}). 

Because $r < R$ across the integration domain, the bracketed term is strictly 
negative, mathematically demonstrating that the spherically enclosed 3D light 
fraction is always smaller than the projected 2D counterpart ($F_{\rm 3D}(r) 
< F_{\rm 2D}(r)$) due to the removal of foreground and background line-of-sight 
light within the projected cylinder on the plane of the sky but outside of the central sphere. Equation~(\ref{eq:F3D_single_integral}) provides a practical numerical 
route to computing 3D light fractions, bypassing the need to compute the 
spatial density $\nu(r)$ first.

The fraction of light enclosed within spheres of radius $r = R_5 \equiv 
R_{-2}$ is reported in column~2 of Table~\ref{tab:light_fractions} for 
different values of $n$. The 3D enclosed fraction reaches a minimum of 
$42.76$ per cent at $n = 0.5$ (the Gaussian case) and increases in both 
directions: rising to $47.56$ per cent at $n = 10$ as the profile becomes 
more centrally concentrated, and rising to $43.04$ per cent at $n = 0.25$. 
The upturn below $n = 0.5$ reflects a qualitative change in the deprojected 
density profile: for $n < 0.5$, the Abel deprojection of the 
S\'ersic profile yields a $\nu(r)$ that increases outward from the centre, 
peaking at $r \approx 0.90\,R_{\rm e}$ for $n = 0.25$ before declining. 
This hollow-centred 3D density distribution concentrates light near $r \sim 
R_{\rm e}$ rather than at the origin, raising the fraction of light enclosed 
within the sphere of radius $r = R_{-2} \approx 1.22\,R_{\rm e}$ relative 
to the $n = 0.5$ minimum.
The projection difference $\Delta_{\rm 2D-3D}$---representing the 
fraction of foreground and background line-of-sight light enclosed 
within the projected cylinder of radius $R_{-2}$ but residing outside 
the sphere of radius $r = R_{-2}$---is given by $F_{\rm 2D}(R_{-2}) - 
F_{\rm 3D}(R_{-2})$, i.e.\ the magnitude of the correction integral in 
Equation~(\ref{eq:F3D_single_integral}), and is reported in column~3 of Table~\ref{tab:light_fractions}.

\subsubsection{The internal radius $r_{-3}$}
\label{Sec-r3-exp-R2}

Although the $R^{1/n}$ profile is not a global power law, a long-known mathematical identity of the Abel integral equation for spherical systems states that a 2D power-law intensity profile of local logarithmic slope $-a$ 
deprojects to a 3D luminosity density profile of local logarithmic slope $-(a+1)$ \citep[e.g.,][]{1908AnPar..25F...1V}. 
For the curved $R^{1/n}$ profile, this relation is approximate, becoming increasingly accurate in regions where the profile is locally close to a power law (i.e., where the curvature of $\ln I$ versus $\ln R$ is small). At the specific projected radius $R_{-2} \equiv R_5$, where the local projected logarithmic slope is exactly $-2$ by construction, the Abel deprojection yields the 3D stellar luminosity density slopes reported in Table~\ref{tab:sersic_slopes}.  This internal slope approaches $-3$
asymptotically as $n \rightarrow \infty$, reaching $-2.94$ at $n=10$.  
In the limit $n \to \infty$, the $R^{1/n}$ profile approaches the global 
power law for which the logarithmic 
slope equals $-2$ at every projected radius (see Equation~\ref{Eq_new_slope}, and thus $\mu(R)$ has a slope of 5) and the deprojected intensity profile has a slope of $-3$ everywhere, i.e., at all radii.  
Expanding the exact Abel-deprojection integral (Equation~\ref{eq:exact_deproj_R5_dimensionless}) 
for large $n$ via the Laplace method, it can also be shown that the deprojected profile approaches a pure power law, 
$\nu(r) \propto r^{-3}$. 
Both $R_{-2} \equiv R_5$ and $r_{-3}$ consequently become 
ill-defined as unique scale radii in this limit. Nevertheless, 
evaluating the relevant integrals in closed form yields
\begin{equation}
\frac{r_{-3}}{R_{-2}} \longrightarrow \frac{{\rm e}}{2} \approx 1.3591 \qquad (n \to \infty).
\label{eq:r3_R5_asymptotic}
\end{equation}
This finite ratio should be understood as the asymptotic limit of the
sequence of well-defined radii for finite $n$, rather than as a ratio of
unique radii belonging to the limiting pure power-law profile itself. 

The sequence $r_{-3}/R_{\rm e}$ converges to the finite limit $\approx 1.606$ as $n\to\infty$, 
with $R_5/R_{\rm e} \to e^{1/6} \approx 1.181$ (Equation~\ref{Eq_one-sixth}). 
The identification of the internal radius where the logarithmic slope of the density profile equals $-3$ for any $n$, and the fact that the radius where this 3D slope equals $-3$ lies approximately near the internal half-light radius, was first demonstrated for S\'ersic/Prugniel-Simien models by \citet{2006AJ....132.2701G}.

The ratio $r_{-3}/R_{-2}$, denoted $g(n)$, can be approximated as 
\begin{equation}
    g(n) \approx 1.347 - 0.097 n^{-1} + 0.018 n^{-2}, 
    \label{Eq_approx_g}
\end{equation}
which is accurate to 0.6 per cent for $0.5 \lesssim n \lesssim 10$ (but 5.8 per cent at $n=0.25$). 
Following the power-expansion style of \citet[their equation~21]{1999MNRAS.309..481L}, 
the ratio $r_{-3}/R_{\rm e}$, denoted $f(n)$, can be approximated as 
\begin{equation}
    f(n) \approx 1.594 - 0.104 n^{-1} + 0.021 n^{-2}, 
    \label{Eq_fun}
\end{equation}
which is accurate to better than 0.5 per cent 
for $0.5 \lesssim n \lesssim 10$, but has a maximum error of 5.6 per cent at $n=0.25$.

An inspection of Table~\ref{tab:sersic_slopes} reveals why the projected scale radius $R_{-2}$ is physically and
observationally advantageous relative to the spatial half-light radius
$r_{1/2}$---defining a sphere enclosing 50 per cent of the light---for the purposes considered here. First, $r_{1/2}$ is a 
3D deprojected quantity that cannot be directly measured on 
the plane of the sky, whereas $R_{-2}$ is a directly observable 2D 
coordinate identifiable from the light profile without deprojection 
(Equation~\ref{eq:Rstar_def}).
Second, while the deprojected stellar density slope at $r=r_{1/2}$ is close 
to $-3$ for high-concentration galaxies ($n\gtrsim 3$), the approximation 
breaks down noticeably for low-concentration systems. For dwarfs and 
UDGs with $n\lesssim 1$, the internal density slope at the 
half-light radius is roughly $-2.4$ to $-2.5$. Assuming $r_{1/2} \approx r_{-3}$ 
therefore introduces a systematic offset precisely in the dwarf regime 
where dark matter mass modelling is most frequently applied. In contrast, 
mapping the directly observable $R_{-2}$ to the exact $r_{-3}$ radius will 
(Table~\ref{tab:sersic_slopes}) provide a robust correspondence 
across the range of S\'ersic indices considered here.
The  fraction of light enclosed within spheres of radius $r=r_{-3}$ is reported in Table~\ref{tab:light_fractions} for different values of $n$ and shown in Figure~\ref{Fig_Sersic_fractions}. 
For completeness, and to quantify how much light lies within the projected
cylinder of radius $R=r_{-3}$ but outside the corresponding sphere, Column~5
of Table~\ref{tab:light_fractions} provides this information.  
The value of this measure will become more relevant in  Section~\ref{Sec_massive}, dealing with dynamical masses.

\subsection{Validity of the Jeans framework for \texorpdfstring{$n < 1.0$}{n1}}

For $n=0.5$, the projected profile is Gaussian and the exact deprojection 
is also Gaussian; both, therefore, have a vanishing central logarithmic slope.  However, as noted earlier, 
for S\'ersic profiles with $n<0.5$, the exact deprojected stellar 
density develops a central depression near the origin (while the projected intensity profile itself 
remains centrally flat or peaked), meaning the spatial density increases with radius 
over the inner regions \citep[e.g.,][]{2019A&A...626A.110B}. Under the additional 
assumptions of spherical symmetry and velocity isotropy, this central 
depression prevents the system from being physically supported: the resulting 
isotropic phase-space distribution function $f(\mathcal{E})$ becomes negative 
near the centre \citep{2019A&A...626A.110B}. 

(Note: The commonly used \citet{1997A&A...321..111P} approximation
(Equation~\ref{eq:PS97_deproj_R5}) does not reproduce this central behaviour.
Across the range $0.55 \lesssim n < 1$, where the true deprojected S\'ersic
density is finite at the origin, the Prugniel-Simien approximation forces an infinite central
cusp ($\alpha(n)>0$, see Equation~\ref{eq:alpha_PS97}). For the hollow-cored profiles with $n<0.5$, the
approximation forces the central density to drop to zero as a power law
rather than reproducing the exact deprojection’s gentler (approximately
parabolic) rise from the centre.)

The central density depression is a mathematical consequence of the exact spherical deprojection. Its interpretation as a physical equilibrium stellar system becomes problematic under the additional assumptions of sphericity, isotropy, and mass following light. However, real ETGs with 
$n\lesssim 1$ are typically (i) flattened (oblate, triaxial, or disc-like\footnote{Disc-like systems may be either rotationally supported or largely pressure-supported (with little net rotation).}), (ii) rotationally supported, or (iii) stabilized by tangential orbital anisotropy ($\beta<0$). Any of these breaks the assumptions that render the hollow-cored isotropic model unphysical. As such, 
dark matter is not required to explain the existence of such profiles, although 
a centrally concentrated dark-matter halo can help maintain physical consistency 
for a nearly spherical, low-$n$ stellar distribution.
In addition, under the same isotropic, mass-follows-light assumptions, it is noted that the radial velocity-dispersion profile declines toward the centre for $n<1$ \citep{1991A&A...249...99C,2002A&A...386..149B}. This 
kinematic behaviour is a second, independent indication that the classical 
isotropic Jeans framework becomes unrealistic in the low-concentration regime.

In passing, an additional caveat is noted: at mid-values of $n$, ordinary ETGs are frequently merger-built composites of a spheroid plus one or more thickened stellar discs that a single S\'ersic fit blends into a single $n$ \citep{2024MNRAS.535..299G}.  The single, relaxed, one-component system implicitly assumed by the isotropic Jeans treatment \citep[Section~\ref{Sec_Bertin}, ][]{1991A&A...249...99C} can therefore be an oversimplification across a wider range of $n$, not only for $n<1$.

\subsection{Re-parameterizing the Prugniel \& Simien deprojection approximation}
\label{Sec_PS}

Following \citet{1997A&A...321..111P}, 
but using the new parameters introduced here, 
the deprojected $R^{1/n}$ luminosity-density profile can be approximated with the following expression
\begin{equation}
\nu(r) \approx \nu_5 \left(\frac{r}{R_5}\right)^{-\alpha(n)} 
\exp\!\left\{ -2n \left[ \left(\frac{r}{R_5}\right)^{1/n} - 1 
\right] \right\},
\label{eq:PS97_deproj_R5}
\end{equation}
where $\nu_5 \equiv \nu(R_5)$ is the luminosity density at 
$r = R_5$, and 
\begin{equation}
\alpha(n) \approx 1 - 0.6097n^{-1} + 0.05463n^{-2}
\label{eq:alpha_PS97}
\end{equation}
\citep[][their equation~19; see also \citealp{1997A&A...321..111P}, their equation~B7]{1999MNRAS.309..481L}.  This form is 
free of both $b_n$ and $R_{\rm e}$, with the $R_5$ scale radius  
replacing $R_{\rm e}$ and $2n$ replacing $b_n$ in the standard 
Prugniel-Simien parameterization, as follows directly from the 
replacements $b_n \to 2n$ and $R_{\rm e} \to R_5$ in 
Equation~(\ref{eq:Sersic_new_I2}).
Following \citet{2006AJ....132.2701G}, the spatial luminosity density at the $R_5$ scale radius, 
$\nu_5 \equiv \nu(R_5)$, can be expressed exactly in terms of the $R^{1/n}$ introduced here as
parameters 
\begin{equation}
\nu_5 = \frac{I_5 (2n)^{n[1-\alpha(n)]} \Gamma(2n)}{2 R_5 \Gamma(n[3-\alpha(n)])}.
\label{eq:nu5_exact}
\end{equation}

The corresponding spatial mass density  
$\rho_5 \equiv \rho(R_5)$ is defined by incorporating the global mass-to-light 
ratio ($M/L$):
\begin{equation}
\rho_5 = (M/L) \nu_5
\label{eq:rho5_exact}
\end{equation}
Applying these parameter conversions to this internal profile yields 
the closed-form, ($b_n, R_{\rm e}$)-free internal-luminosity distribution: 
\begin{equation}
L(r) = 4\pi n R_5^3 \nu_5 e^{2n} (2n)^{[\alpha(n)-3]n} \gamma\left( [3-\alpha(n)]n,\; z \right),
\label{eq:mass_profile_R5}
\end{equation}
where $z = 2n (r/R_5)^{1/n}$ represents the dimensionless integration limit, 
and $\gamma(a, z)$ is the lower incomplete gamma function.
A variant of this analytical integration for the enclosed spatial (internal) luminosity 
distribution was derived by \citet{1999MNRAS.309..481L}.

While the exact deprojected spatial radius $r_{-3}$ must be computed 
numerically via the Abel integral (Equation~\ref{eq:exact_deproj} or \ref{eq:exact_deproj_R5_dimensionless}), an elegant 
analytical approximation follows from the logarithmic slope of the Prugniel--Simien approximation 
in Equation~\ref{eq:PS97_deproj_R5}:
\begin{equation}
\frac{\mathrm{d}\ln\nu}{\mathrm{d}\ln r} = -\alpha(n) - 2 \left(\frac{r}{R_5}\right)^{1/n}.
\label{eq:PS97_slope}
\end{equation}
Setting this approximate local gradient equal to $-3$ at $r = r_{-3}$ gives 
\begin{equation}
\frac{r_{-3}}{R_5} = \left[ \frac{3 - \alpha(n)}{2} \right]^n \approx \left[ 1 + \frac{0.3049}{n} - \frac{0.02732}{n^2} \right]^n.
\label{eq:r_minus3_ratio_analytical}
\end{equation}
This deprojection-derived relation is in excellent agreement with the 
direct, empirical polynomial fit for $g(n)$ presented in 
Equation~(\ref{Eq_approx_g}).\footnote{At $n=0.25$, where the numbers are most discrepant, Equation~\ref{Eq_approx_g} is too high by $\sim$6 per cent, while Equation~\ref{eq:r_minus3_ratio_analytical} is too low by just $\sim$2 per cent.}

\section{Masses}

\subsection{A refined \texorpdfstring{$r_{-3}$}{rrr-3} Wolf-type dynamical mass estimator involving \texorpdfstring{$R_5$}{Rrr5}}
\label{Sec_massive}

In a remarkable instance of concurrent discovery, both \citet{2010MNRAS.404.1165C} and \citet{2010MNRAS.406.1220W} independently demonstrated that the mass-anisotropy degeneracy in the spherical Jeans equation is minimized near $r=r_{-3}$.
\citet{2010MNRAS.406.1220W} showed that the mass enclosed within the internal (3D) radius $r_{-3}$ (where the deprojected density slope is $-3$) is minimally sensitive to the velocity dispersion anisotropy $\beta$, provided that the luminosity-weighted line-of-sight velocity dispersion profile is relatively flat near and beyond the projected radius corresponding to $r_{-3}$. 
Their primary mass estimator 
\begin{equation}
M(<r_{-3}) = 3\, G^{-1} \langle \sigma_{\rm los}^2 \rangle r_{-3},
\label{eq:wolf_r3}
\end{equation}
requires $\langle\sigma_{\rm los}^2\rangle$ to be the global luminosity-weighted second moment, while an aperture measurement is used as an approximation once the luminosity-weighted dispersion profile is sufficiently flat.
In practice, \citet{2010MNRAS.406.1220W} suggested that the global luminosity-weighted second moment can often be approximated by an aperture measurement extending to approximately $R_{\rm e}$.  
To express this mass estimator in terms of familiar projected observables, \citet{2010MNRAS.406.1220W} adopted the approximation $r_{-3} \approx r_{1/2}$ \citep{2006AJ....132.2701G} and $r_{1/2} \approx 1.33\,R_{\rm e}$ \citep{1991A&A...249...99C}, yielding their widely used Wolf mass estimator:
\begin{equation}
M_{1/2} \equiv M(<r_{1/2}) \simeq 3\, G^{-1} \langle \sigma_{\rm los}^2 \rangle r_{1/2} \simeq 4\, G^{-1} \langle \sigma_{\rm los}^2 \rangle \,R_{\rm e}.
\label{eq:wolf_estimator}
\end{equation}
The (dynamical mass)-to-light ratio within a sphere of radius $r_{1/2}$ is thus:
\begin{equation}
\left(\frac{M_{\rm dyn}}{L}\right)_{1/2} = \frac{M_{1/2}}{0.5L_{\rm tot}}.
\label{Eq:ML_half}
\end{equation}

However, as shown in Table~\ref{tab:sersic_slopes}, and noted by \citet[][their Table~B1]{2010MNRAS.406.1220W}, $r_{-3}$ is systematically $13$ to $18$ per cent larger than $r_{1/2}$ for S\'ersic profiles with $n=0.5$ to 8 (e.g., $r_{-3} \approx 1.57\,R_{\rm e}$ for $n=4$). Substituting $r_{1/2}$ directly for $r_{-3}$ without adjusting the coefficient in Equation~(\ref{eq:wolf_estimator}) thus introduces a systematic underestimation of the mass at the true anisotropy-insensitive radius $r_{-3}$. Indeed, for an $n=4$ profile, $M(<r_{-3})_{n=4} \approx 4.7\,G^{-1}\langle \sigma_{\rm los}^2 \rangle \,R_{\rm e}$. 
Moreover, the shape of the deprojected luminosity profile at $r=r_{1/2}$ increasingly deviates from a value of $-3$ as $n$ decreases (Table~\ref{tab:sersic_slopes}).

A more precise mass estimator than $M_{1/2}$ is obtained here by retaining the 
exact anisotropy-insensitive radius $r_{-3}$ and expressing it through the 
observable scale $R_5$. The refined Wolf-type mass estimator is defined as
\begin{equation}
M_{-3} \equiv M(<r_{-3}) = 3\, G^{-1} \langle \sigma_{\rm los}^2 \rangle\, g(n)\, R_5,
\label{eq:mass_R5}
\end{equation}
where $g(n)\equiv r_{-3}/R_5$ is the weakly $n$-dependent galaxy size transformation function given in 
Table~\ref{tab:sersic_slopes}, along with other scale radii transformations. 
By discarding the spatial half-light approximation of \citet{2010MNRAS.406.1220W} (and the power-law approximation of \citet{2010MNRAS.404.1165C}, discussed in Section~\ref{Sec_Churros}), an $n$-dependent 
Wolf-type mass estimator is obtained that consistently relates the projected scale $R_{-2}$ to the deprojected anisotropy-insensitive radius $r_{-3}$ for finite S\'ersic indices.

The approximation to $g(n)$, given by Equation~\ref{Eq_approx_g}, can be used in Equation~\ref{eq:mass_R5}.  
Given $g(n)$ is remarkably stable against different galaxy concentrations, remaining close to $4/3 \approx 1.33$ for $n \gtrsim 2$, one has 
\begin{equation}
M_{-3} \approx 4\, G^{-1} \langle \sigma_{\rm los}^2 \rangle R_5, \,\, {\rm for}\,\, n\gtrsim 2.
\label{eq:mass_R5_approx}
\end{equation}
Expressing this in terms of the standard effective half-light radius via the relation $R_5/R_{\rm e} = (2n/b_n)^n \approx 1.18$ (Equation~\ref{Eq_R5} and Table~\ref{Tab_R5_ratios}) yields:
\begin{equation}
M_{-3} \approx 4.72\, G^{-1} \langle \sigma_{\rm los}^2 \rangle R_{\rm e}, \,\, {\rm for}\,\, n\gtrsim 2.
\label{eq:mass_Re_exact}
\end{equation}

For greater accuracy, in terms of the standard effective half-light radius, substituting
$R_5=(2n/b_n)^n\,R_{\rm e}$ into Equation~(\ref{eq:mass_R5}) gives
\begin{equation}
M_{-3} = 3\,G^{-1}\langle\sigma_{\rm los}^2\rangle\,f(n)\,R_{\rm e},
\label{eq:mass_Re_exact-2}
\end{equation}
where the composite function $f(n)$---representing the fundamental 
scaling function (Table~\ref{tab:sersic_slopes})---is the product of 
the galaxy size transformation function $g(n)$ (Table~\ref{tab:sersic_slopes}) 
and the ratio of scale radii $R_5/R_{\rm e}$ (Table~\ref{Tab_R5_ratios}):
\begin{equation}
f(n)\equiv\frac{r_{-3}}{R_{\rm e}}
        = \left( \frac{r_{-3}}{R_5}\right)  \left( \frac{R_5}{R_{\rm e}}\right) 
          =g(n)\left(\frac{2n}{b_n}\right)^n.
\end{equation}
For an $n=4$ profile, this evaluates to $f(4) \approx 1.57$, recovering the $n=4$ mass estimator $M_{-3} \approx 4.7\, G^{-1} \langle \sigma_{\rm los}^2 \rangle R_{\rm e}$.
Other values of $f(n)$ can be obtained from the approximation given by Equation~\ref{Eq_fun}.

\subsection{Dynamical mass-to-light ratios}
\label{Sec_mottle}

The true spatial (not projected) (dynamical mass)-to-light ratio within the
deprojected sphere of radius $r=r_{-3}$ is defined by dividing the 3D spatial 
mass $M(<r_{-3})$ by the 3D spatial deprojected luminosity $L_{\rm 3D}(<r_{-3})$:
\begin{equation}
\left(\frac{M_{\rm dyn}}{L}\right)_{-3} \equiv \frac{M(<r_{-3})}{L_{\rm 3D}(<r_{-3})}.
\label{Eq_swallowable}
\end{equation}
Substituting the mass estimator $M(<r_{-3}) = 3\, G^{-1} \langle \sigma_{\rm 
los}^2 \rangle\, g(n) R_5$ (Equation~\ref{eq:mass_R5}) and the enclosed 
spatial deprojected luminosity $L_{\rm 3D}(<r_{-3}) = L_{\rm tot} \times 
\left[L(<r=r_{-3})\% / 100\right]$ (with the exact fractions given in 
Column~4 of Table~\ref{tab:light_fractions}) yields:
\begin{equation}
\left(\frac{M_{\rm dyn}}{L}\right)_{-3} = \frac{3\, G^{-1} \langle \sigma_{\rm los}^2 
\rangle\, g(n) R_5}{L_{\rm tot} \left[L_{\rm 3D}(<r_{-3})/
L_{\rm 
tot}\right]} = \frac{3}{G} \, h(n) \frac{R_5 \langle \sigma_{\rm los}^2 
\rangle}{L_{\rm tot}},
\label{Eq_mouthfull}
\end{equation}
where the handy dimensionless function $h(n)$ (Table~\ref{tab:fn_vs_KV}) is defined as:
\begin{equation}
h(n) \equiv \frac{g(n)}{L_{\rm 3D}(<r_{-3})/L_{\rm tot}}.
\label{eq:hn_definition}
\end{equation}
This formulation allows users to calculate the spatial mass-to-light 
ratio inside the $r_{-3}$ sphere directly from the total luminosity $L_{\rm 
tot}$, sparing them from having to integrate the spatial density profile to 
find $L_{\rm 3D}(<r_{-3})$ individually.

Following, again, the power-expansion style of \citet[their equation~21]{1999MNRAS.309..481L}, 
this new spatial mass-to-light scaling parameter can be approximated 
across the range ($0.25 \le n \le 10$) as:
\begin{equation}
h(n) \approx 2.565 - 0.448 n^{-1} + 0.066 n^{-2},
\label{eq:hn_approximation}
\end{equation}
which is accurate to better than $1.4$ per cent across almost all $n$, with the exception of a maximum relative error of $\approx$4 per cent at $n = 0.5$.

Combined with $F_{\rm 3D}(r_{-3}) \to 1/2$ as $n$ tends to 
infinity\footnote{The limit \( F_{\rm 3D}(r_{-3})\to 1/2 \) follows from an asymptotic 
expansion of Equation~(\ref{eq:F3D_single_integral}).  At \( R=R_5 \) 
one has \( x=2n \), so the projected fraction is the incomplete-gamma 
ratio \( \gamma(2n,2n)/\Gamma(2n) \), which is known to tend to \( 1/2 \) 
as \( n\to\infty \).  The geometric 2D--3D projection correction 
vanishes in the same limit, 
yielding \( F_{\rm 3D}(r_{-3})\to 1/2 \).}, 
$h(n)$ has the exact asymptotic limit
\begin{equation}
h(n) \longrightarrow \frac{{\rm e}/2}{1/2} = {\rm e} \approx 2.7183 \qquad (n \to \infty),
\label{eq:h_infty_exact}
\end{equation}
an elegant closed-form result connecting the dynamical mass-to-light scaling to Euler's number, $\rm e$ (see Appendix~\ref{Appdx_Sec_r3_asymptotic})

For applications where it is desirable to express the mass-to-light ratio 
directly in terms of the projected effective half-light radius $R_{\rm e}$, 
substituting the scaling relation $M(<r_{-3}) = 3\, G^{-1} \langle \sigma_{\rm 
los}^2 \rangle\, f(n) R_{\rm e}$ (Equation~\ref{eq:mass_Re_exact-2}) into 
Equation~(\ref{Eq_swallowable}) yields:
\begin{equation}
\left(\frac{M_{\rm dyn}}{L}\right)_{-3} = \frac{3}{G} \left[ \frac{f(n)}{F_{\rm 3D}(r_{-3})} \right] \frac{R_{\rm e} \langle \sigma_{\rm los}^2 \rangle}{L_{\rm total}},
\label{eq:ML_Re_exact}
\end{equation}
where $F_{\rm 3D}(r_{-3}) \equiv L_{\rm 3D}(<r_{-3})/L_{\rm total}$ is the 
spatial light fraction from column~4 of Table~\ref{tab:light_fractions}. 

Furthermore, this formulation allows a direct comparison with the 
spatial half-light mass-to-light ratio derived from the Wolf 
 mass estimator (Equation~\ref{Eq:ML_half}). Under the assumption of flat velocity dispersion profiles, 
the ratio of the two spatial mass-to-light estimators is given by the quotient 
\begin{equation}
Q(n) \equiv \frac{(M_{\rm dyn}/L)_{-3}}{(M_{\rm dyn}/L)_{1/2}} = \left( \frac{r_{-3}}{r_{1/2}} \right) \frac{0.5}{F_{\rm 3D}(r_{-3})}.
\label{eq:Q_n_ratio}
\end{equation}
By anchoring the leading term of the expansion to the exact asymptotic limit 
of $Q(\infty) = 1.185$, this systematic, $n$-dependent offset can be 
conveniently evaluated across the 
sequence ($0.25 \le n \le 10$) using the compact expression:
\begin{equation}
Q(n) \approx 1.185 - 0.260\, n^{-1/2} + 0.053\, n^{-1},
\label{eq:Q_n_fit}
\end{equation}
which is accurate to better than $0.6$ per cent down to $n = 0.25$ and better 
than $0.2$ per cent across the range $0.5 \le n \le 5$. 

Because of S\'ersic structural non-homology, $Q(n)$ is a smoothly increasing 
function of the S\'ersic index. For high-concentration ETGs with 
$n > 1.5$, the ratio is greater than unity (e.g., $Q(4) \approx 1.070$), reflecting 
the fact that the larger volume of the $r_{-3}$ sphere encloses very little 
additional light beyond the highly concentrated core. Conversely, for 
low-concentration dwarf and UDGs with $n < 1.5$, the 
ratio is smaller than unity (e.g., $Q(0.5) \approx 0.925$ and $Q(0.25) \approx 
0.871$; see Table~\ref{tab:fn_vs_KV}). In this low-$n$ regime, the rapid growth in enclosed spatial light 
between $r_{1/2}$ and $r_{-3}$ outpaces the linear radius growth. Evaluating 
the mass-to-light ratio inside the larger, dynamically robust $r_{-3}$ volume 
therefore yields even lower, more conservative values for low-$n$ dwarf systems than 
those obtained at the half-light radius.

\subsection{Some prior dynamical mass estimators (assuming a dark matter halo)}
\label{Sec_prior_mass}

\subsubsection{Walker et al.\ (2009)}

\citet{2009ApJ...704.1274W} derived a simple analytic mass estimator evaluated at the projected effective half-light radius $R_{\rm e}$. Under the simplifying assumptions of velocity isotropy ($\beta=0$), a flat velocity dispersion profile, and a \citet{1911MNRAS..71..460P} stellar density profile, they solved the spherical Jeans equation to show that the 3D spatial mass enclosed within the sphere of radius $r = R_{\rm e}$ is given by:
\begin{equation}
M(<R_{\rm e}) = 2.5\, G^{-1} \langle \sigma_{\rm los}^2 \rangle R_{\rm e}.
\label{eq:walker_sub}
\end{equation}
While robust and widely applied to dwarf spheroidal galaxies, this estimator is structurally tied to the specific assumption and shape of the Plummer profile.

\subsubsection{Wolf et al.\ (2010)}

As noted in Section~\ref{Sec_massive}, 
\citet{2010MNRAS.406.1220W} approached the mass-anisotropy degeneracy from a 3D deprojected perspective, showing that the integrated total mass (stars plus dark matter) is most robustly constrained at the spatial radius $r_{-3}$ where the deprojected tracer density slope is exactly $-3$. This led to their spatial half-light mass estimator $M_{1/2} \simeq 4\, G^{-1} \langle \sigma_{\rm los}^2 \rangle R_{\rm e}$ (Equation~\ref{eq:wolf_estimator}). 

As shown in Table~\ref{tab:sersic_slopes}, the deprojected radius $r_{-3}$ is systematically $11$ to $18$ per cent larger than $r_{1/2}$ across the S\'ersic sequence ($n = 0.25$--$10$). By using the deprojected scaling function $g(n) \equiv r_{-3}/R_5$ (Equation~\ref{eq:mass_R5}) rather than a constant approximation, the refined Wolf-type mass estimator $M_{-3}$ removes the systematic
concentration-dependent offset introduced by the half-light approximation, offering a more precise and physically consistent mass determination across the observed range of S\'ersic indices. 
The $M_{-3}$ estimator, evaluated at the dynamically preferred $r_{-3} \approx (1.27$--$1.36)\,R_5 \approx (1.51$--$1.61)\, R_{\rm e}$ (for $1 < n < \infty $, Table~\ref{tab:sersic_slopes}), refines the mass estimate by targeting the radius where the mass-anisotropy degeneracy is minimized. 
When expressed in terms of the projected scale $R_5$, the refined
estimator takes the convenient form
$M_{-3}\approx4\,G^{-1}\langle\sigma_{\rm los}^2\rangle R_5$ for $n\gtrsim2$
(converging to $\approx4.08\,G^{-1}\langle\sigma_{\rm los}^2\rangle R_5$ as $n\to\infty$).
However, when mapped back to the standard effective radius $R_{\rm e}$,
the true coefficient is $3f(n)\approx4.7$--$4.8$, demonstrating that the
classic Wolf prefactor of $4$ applied to $R_{\rm e}$ systematically
underestimates the mass enclosed within $r_{-3}$ by $\approx15$--$18$ per cent.

The present work provides a unified treatment that determines the
anisotropy-insensitive radius $r_{-3}$ from the observable scale
$R_{-2}$ through an exact numerical deprojection and an accurate,
weakly $n$-dependent transformation function $g(n)$ for finite
S\'ersic indices. It additionally quantifies the difference between $(M_{\rm dyn}/L)_{1/2}$ and the preferable value $(M_{\rm dyn}/L)_{-3}$.

\subsubsection{Churazov et al.\ (2010)}
\label{Sec_Churros}

\citet{2010MNRAS.404.1165C} developed several robust estimators for 
the circular velocity \(v_c\) of a spherical galaxy whose gravitational 
potential is approximately isothermal (\(\phi=v_c^2\log r\)). In such a 
potential the circular speed is constant with radius and provides a 
direct measure of the enclosed mass via \(v_c^2=GM(r)/r\).  
They identified two related ``sweet spot'' radii at which the inferred circular speed is relatively insensitive to the stellar orbital anisotropy. Their primary sweet spot, $R_{\rm s}\approx0.5\,R_{\rm e}$, is defined dynamically by the condition $\sigma_{\rm circ}(R_{\rm s})=\sigma_{\rm rad}(R_{\rm s})$, where the predicted line-of-sight velocity dispersions for purely circular and purely radial stellar orbits coincide. This radius depends only weakly on the S\'ersic index and provides a practical location at which the observed velocity dispersion is a robust proxy for the circular speed.  

A second, local estimator was obtained by expressing the Jeans equation in terms of the local, negative, logarithmic intensity profile slope $\alpha\equiv -\mathrm{d}\ln I/\mathrm{d}\ln R$.  Neglecting higher-order derivatives of the velocity-dispersion profile,
their expressions for isotropic, radial, and circular orbits reduce to
$v_{\rm c}\approx\sqrt{3}\,\sigma_{\rm los}$ at the projected radius where
$\alpha=2$, corresponding to $R_{-2}$ (also denoted $R_5$ in the present work). Their result is only exact/applicable for a projected power-law intensity profile with $I(R)\propto R^{-2}$, which deprojects via the Abel transform to $\nu(r)\propto r^{-3}$. 

However, 
for real galaxies with $R^{1/n}$ light profiles and finite values of $n$, the projected and intrinsic logarithmic slopes 
define two distinct characteristic scales. Using the (infinite-$n$) approximation 
$b_n\approx2n$, \citet{2010MNRAS.404.1165C} obtained 
$\alpha(R)\approx2(R/R_{\rm e})^{1/n}$, such that the projected radius 
satisfying $\alpha=2$ reduces to $R\approx R_{\rm e}$. The exact analytical 
expression derived here---$R_{-2} \equiv R_5 = (2n/b_n)^n \, R_{\rm e}$ (Equation~\ref{Eq_R5})---shows that the projected radius where the surface-brightness slope equals 
$-2$ is actually $\approx18$--$22$ per cent larger than $R_{\rm e}$ over the 
range of S\'ersic indices observed in galaxies. Moreover, the internal radius satisfying 
$\mathrm{d}\ln\nu/\mathrm{d}\ln r=-3$ is still further out.  
Consequently, the characteristic projected radius $R_{-2}$ and intrinsic 
radius $r_{-3}$ represent distinct physical scales for finite S\'ersic 
profiles, although this distinction disappears in the limiting case of a 
pure $R^{-2}$ power law intensity profile, for which the projected and 
deprojected logarithmic slopes are $-2$ and $-3$, respectively, at all 
radii. 

This second projected sweet spot identified by 
\citet{2010MNRAS.404.1165C} and the intrinsic sweet spot identified by  
\citet{2010MNRAS.406.1220W} 
are mathematically connected through the tracer-density profile, and correspond to distinct physical radii for finite-$n$. 
The relations derived in this work provide the analytical mapping  
between these scales, thereby connecting the observable quantity 
$R_{-2}\equiv R_5$ to the anisotropy-insensitive radius $r_{-3}$ for ETGs of 
arbitrary S\'ersic indices.

\subsubsection{Lyskova et al.\ (2015)}
\label{Lyskova}

\citet{2015MNRAS.450.3442L} 
systematically compared two simple mass estimators 
for elliptical galaxies across analytical S\'ersic models, cosmological simulations, 
and real early-type galaxies previously modelled with Schwarzschild 
orbit-superposition techniques. One estimator 
\citep[from][]{2010MNRAS.404.1165C} 
is local: it evaluates the circular speed from the velocity dispersion at the 
projected radius $R_2$ where the surface-brightness profile declines as $R^{-2}$ 
(identifying the same projected landmark as $R_5\equiv R_{-2}$). The other 
\citep[from][]{2010MNRAS.406.1220W} 
is global: it evaluates the mass enclosed within the deprojected three-dimensional half-light radius $r_{1/2}$ using the global luminosity-weighted velocity dispersion. 
\citet{2015MNRAS.450.3442L} 
found that both estimators recover 
nearly unbiased circular speeds with modest rms scatter ($\lesssim10$ per cent), 
with the local $R_2$ estimator exhibiting slightly lower scatter than the 
global $r_{1/2}$ estimator. The present work complements their numerical tests 
by providing the exact analytical mapping between the observable projected 
gradient scale $R_5\equiv R_{-2}$ and the intrinsic anisotropy-insensitive radius 
$r_{-3}$ across the entire S\'ersic sequence ($0.25 \le n \le 10$), clarifying the 
structural and volume offsets that arise when $r_{1/2}$ is used as a proxy for $r_{-3}$.

\begin{table}
\centering
\caption{
Comparison of the mass-estimator coefficient $3f(n)$ for stellar systems
embedded in dark-matter halos and having an approximately flat,
luminosity-weighted outer-aperture velocity-dispersion profile, with the
reduced mass-follows-light virial coefficient
$K_{V,\rm red}(n)$ (based on $R_{\rm e}/10$ velocity dispersion apertures) for stellar systems without dark matter.
The former enters the Wolf-type estimate
$M(<r_{-3})=3f(n)\,\sigma^2R_{\rm e}/G$, whereas the latter is obtained
by reducing the total-mass virial coefficient $K_V(n)$ to that matching the mass enclosed
within $r_{-3}$, and enters
$M(<r_{-3})=K_{V,\rm red}(n)\,\sigma^2R_{\rm e}/G$.
The two estimators use different velocity-dispersion apertures and make
different physical assumptions regarding the relation between light and
mass (see Section~\ref{Sec_prior_mass}).
The function $h(n)$ from Equation~\ref{eq:hn_definition} is tabulated in Column~3, while 
values of $g(n) \equiv r_{-3}/R_{-2} \equiv r_{-3}/R_{5}$ are provided in Table~\ref{tab:sersic_slopes}. 
Column~7 gives $Q(n) = (M_{\rm dyn}/L)_{-3} \,/ \, (M_{\rm dyn}/L)_{1/2}$ (Equation~\ref{eq:Q_n_ratio}). 
Values related to $K_V(n)$ are restricted to $0.5 \le n \le 10$; entries for S\'ersic indices $n < 0.5$ are omitted 
because isotropic mass-follows-light solutions become unphysical in this regime.
}
\label{tab:fn_vs_KV}
\begin{tabular}{lcccccc}
\hline
$n$    & $3f(n)$ & $h(n)$ & $K_V(n)$ & $K_{V,\,\rm red}$ & $3f/K_{V,\,\rm red}$ &  $Q(n)$ \\
(1)    &   (2)   &  (3)   &   (4)    &    (5)            &  (6)                 &  (7)   \\
\hline
$0.25$ & $4.302$ & $1.825$ & $\dots$ & $\dots$           & $\dots$              &  0.871 \\
$0.5$ & $4.413$ & $2.014$ & $6.12$  & 3.72               & 1.19                 &  0.925 \\
$0.6$ & $4.445$ & $2.060$ & $6.53$ & $3.92$            & $1.13$              & $0.941$ \\
$0.7$ & $4.472$ & $2.098$ & $6.91$ & $4.11$            & $1.09$              & $0.954$ \\
$0.8$ & $4.496$ & $2.131$ & $7.24$ & $4.27$            & $1.05$              & $0.964$ \\
$0.9$ & $4.516$ & $2.158$ & $7.53$ & $4.40$            & $1.03$              & $0.973$ \\
$1$   & $4.534$ & $2.182$ & $7.75$ & $4.50$            & $1.01$              &  0.978 \\
$1.5$ & $4.599$ & $2.269$ & $8.06$ & $4.58$            & $1.00$              & $1.011$ \\
$2$   & $4.640$ & 2.325 & $7.53$  & 4.22                & $1.10$               &  1.029 \\
$3$   & $4.688$ & 2.395 & $5.97$  & 3.29                & $1.42$               &  1.054 \\
$4$   & $4.715$ & 2.437 & $4.70$  & 2.56                & $1.84$               &  1.070 \\
$5$   & $4.733$ & 2.468 & $3.79$  & 2.05                & $2.31$               &  1.082 \\
$6$   & $4.746$ & 2.490 & $3.13$  & 1.68                & $2.83$               &  1.090 \\
$7$   & $4.755$ & 2.507 & $2.63$  & 1.41                & $3.37$               &  1.097 \\
$8$   & $4.762$ & 2.521 & $2.25$  & 1.20                & $3.97$               &  1.102 \\
$9$   & $4.768$ & 2.533 & $1.95$  & 1.03                & $4.63$               &  1.107 \\
$10$  & $4.772$ & 2.543 & $1.72$  & 0.91                & $5.24$               &  1.112 \\
$\rightarrow \infty$ & $4.818$ & $2.718$&  $\dots$  & $\dots$ & $\dots$        & 1.185 \\
\hline
\end{tabular}
\end{table}

\subsection{Some prior and new dynamical mass estimators (in the absence of dark matter)}

\subsubsection{Bertin et al.\ (2002)}
\label{Sec_Bertin}

In the absence of dark matter, and thus the absence of extended halos and roughly flat luminosity-weighted aperture velocity dispersion profiles, the Virial theorem dictates that the velocity dispersion term in the scaling $M \propto \sigma^2 R$ is the luminosity-weighted aperture velocity dispersion within an infinite aperture and the radial term is the gravitational (harmonic mean) radius, $R_g$. 
Such an arrangement applies to stellar systems like globular clusters and perhaps dynamically relaxed tidal dwarf galaxies. 
For such systems, it is therefore necessary to have an alternative expression to that given by Equation~\ref{eq:mass_R5} (alternatively expressed as Equation~\ref{eq:mass_Re_exact-2}) due to the declining luminosity-weighted aperture velocity dispersion profiles.

\citet{2002A&A...386..149B} adopted a constant mass-to-light ratio with radius, and defined the virial coefficient \(K_V(n)\) through 
\begin{equation}
M_{\rm dyn,tot} = K_V(n) G^{-1} R_{\rm e}\, \langle \sigma_{\rm los}^2 \rangle_{R_{\rm e}/10}.
\label{Eq_mass_KV}
\end{equation}  
This single coefficient \(K_V(n)\) absorbs three distinct factors that 
appear when the classical virial theorem is rewritten in terms of 
observables: (i) the conventional numerical factor 3 that 
accounts for motion in three spatial dimensions (not just along the line-of-sight), (ii) the conversion from the 
small-aperture ($R_{\rm e}/10$) velocity dispersion \(\langle\sigma_{\rm los}^2\rangle_{R_{\rm e}/10}\) 
to the global (infinite-aperture) luminosity-weighted dispersion, and 
(iii) the conversion from \(R_{\rm e}\) to the gravitational 
(harmonic-mean) radius \(R_g\).  
This tripartite decomposition also mirrors the 
earlier, more explicit construction of \citet[][their eqs.~6--10]{1997MNRAS.287..221G}, 
who factored their analogous structural-kinematical coefficient as 
\(K_{\rm SR} = k_{\rm e}/(2 G k_R k_L k_V)\), with separate terms for the 
degree of virialization (\(k_{\rm e}\)), the radius conversion (\(k_R\)), 
the luminosity-structure conversion (\(k_L\)), and the velocity/aperture 
conversion (\(k_V\)).  
\(K_{\rm SR}\) and $K_V$ are functionally identical to the 
\(S_D\) factor tabulated by \citet[][their equations~B14 and B15, and 
Table 4]{1997A&A...321..111P}.  
While \cite{1997A&A...321..111P} obtained their $S_D$ values, for integer values of $n$ from 1 to 10, based on their model approximation of the internal density profile (their equation~B6, itself a generalization of the model from \citet{1987A&A...175....1M} that was designed to approximate the deprojected $R^{1/4}$ model), \citet[][their equation~11, valid for $1 \le n \le 10$]{2002A&A...386..149B} offer $K_V(n)$ values derived from a convenient rational-function representation (their Eq.~11, stated to be accurate 
to $\sim$1 per cent relative to their own numerical solutions). 
Those $K_V(n)$ values differ at the few per cent level from the solutions derived here via the same numerical Abel
inversion of the $R^{1/n}$ model, 
combined with a direct solution of the isotropic Jeans equation, and then a projection onto the plane of the sky followed by an integration over the `sky' to give the aperture velocity dispersion profile, 
following the approach of \citet{1980MNRAS.190..873B}, as extended by \citet{1991A&A...249...99C} and then \citet[][their appendix~B]{1997MNRAS.287..221G}.  This results in the following piecewise fit
\begin{equation} 
K_V(n,0.1R_{\rm e}) \approx 
\begin{cases}
3.155 + 7.221\,n - 2.633\,n^2 & (0.5 \le n \le 1.5), \\[6pt]
-0.804 + 27.35\,n^{-1} - 21.08\,n^{-2} & (1.5 < n \le 10),
\end{cases}
\label{Eq_Kv_new}
\end{equation}
providing an approximation for luminosity-weighted velocity dispersions within apertures of $0.1\,R_{\rm e}$. 
The expression for $0.5 \le n \le 1.5$ matches the entries in Table~\ref{tab:fn_vs_KV} to 0.01 accuracy, 
while the expression for $1.5 < n \le 10$ matches to $\sim$1.6 per cent accuracy. 
The piecewise form captures the initial rise in $K_V(n)$ from the
Gaussian ($n=0.5$) regime characteristic of low-$n$ dwarf and
ultra-diffuse galaxies to a maximum of $K_V\approx8.06$ near
$n\approx1.5$, followed by a monotonic decline toward higher
S\'ersic indices. 
For $n=0.5$, the gravitational radius is
$R_g\approx3.011R_{\rm e}$ (equivalently,
$R_g/R_5\approx2.507$), while
$\sigma_{\rm tot}^2/
\langle\sigma_{\rm los}^2\rangle_{R_{\rm e}/10}=0.678$.
Consequently,
\begin{equation}
3\frac{R_g}{R_{\rm e}}
\frac{\sigma_{\rm tot}^2}
{\langle\sigma_{\rm los}^2\rangle_{R_{\rm e}/10}}
=6.12,
\end{equation}
in agreement with the $K_V(n=0.5)$ value listed in
Table~\ref{tab:fn_vs_KV}.
It should be emphasized that $K_V(n)$ needs to be rederived if using $\langle\sigma_{\rm los}\rangle$ measurements in apertures different from $R_{\rm e}/10$. The aperture velocity dispersion curves first shown in \citet{1997MNRAS.287..221G} reveal the nature of the required adjustments. This is built on in Section~\ref{Sec_GC97_ext}. 

Because the masses derived from Equation~\ref{Eq_mass_KV} refer to a total-mass, comparison with Equation~\ref{eq:mass_Re_exact-2} at the same physical radius requires reducing 
$K_V(n)$ by the fraction of mass that lies inside the sphere of radius
$r_{-3}$. Under the assumption of a constant mass-to-light ratio with
radius, this mass fraction is exactly equal to the deprojected 3D spatial
light fraction $F_{\rm 3D}(r_{-3}) \equiv L_{\rm 3D}(<r_{-3})/L_{\rm total}$
reported in Column~4 of Table~\ref{tab:light_fractions}. This yields the
enclosed-mass coefficient
\begin{equation}
K_{V,\,\rm red}(n) \equiv K_V(n) \times F_{\rm 3D}(r_{-3})
\end{equation}
reported in column~5 of Table~\ref{tab:fn_vs_KV}.

\begin{table*}
\caption{$K_V(n,R_{\rm ap})$ at a range of apertures expressed in units of $R_5$, following
$M_{\rm total} = K_V(n)\,G^{-1}\,R_5\,\langle\sigma_{\rm los}^2\rangle_{\rm aperture}$.
Approximate $R_{\rm e}$-equivalents (since $R_5/R_{\rm e}\approx1.18$--$1.20$ across
this range of $n$; see Column~2) are $R_5/10\!\approx\!R_{\rm e}/8$,
$R_5/5\!\approx\!R_{\rm e}/4$, $R_5/3\!\approx\!0.4R_{\rm e}$,
$R_5/2\!\approx\!0.6R_{\rm e}$, $2R_5/3\!\approx\!0.8R_{\rm e}$, and
$4R_5/5\!\approx\!R_{\rm e}$. The final column gives the infinite-aperture limit,
$K_V(n)=3\,R_g/R_5$ (Column~4), where $R_g$ is the
gravitational (harmonic-mean) radius.
The corresponding reduced coefficients $K_{V,\,\rm red}(n)$ at each aperture
are obtained from $K_{V,\,\rm red}(n)=K_V(n)\,F_{\rm 3D}(r_{-3})$,
where $F_{\rm 3D}(r_{-3})$ is the fraction of the total (3D) 
luminosity enclosed within $r_{-3}$, using the aperture-independent values
tabulated in Table~\ref{tab:light_fractions}. Thus, $K_{V,\,\rm red}$ gives
the mass within $r_{-3}$ corresponding to the total mass-follows-light
virial coefficient $K_V$, for the same aperture velocity dispersion.
}
\label{tab:KV_apertures}
\begin{tabular}{ccccccccccccc}
\hline

$n$ & $R_5/R_{\rm e}$ & $R_g/R_{\rm e}$ & $R_g/R_5$ & \multicolumn{9}{c}{$K_V(n,R_{\rm ap})$}  \\
\multicolumn{4}{c}{}  & $R_5/10$ & $R_5/5$ & $R_5/3$ & $R_5/2$ & $2R_5/3$ & $4R_5/5$ & $R_5$ & $2R_5$ & $\infty$ \\
\hline
0.5 & 1.201 & 3.011 & 2.507 & 5.099 & 5.139 & 5.233 & 5.410 & 5.646 & 5.866 & 6.225 & 7.416 & 7.520  \\
0.6 & 1.198 & 3.058 & 2.552 & 5.448 & 5.447 & 5.495 & 5.629 & 5.828 & 6.019 & 6.335 & 7.476 & 7.657  \\
0.7 & 1.196 & 3.097 & 2.590 & 5.763 & 5.702 & 5.697 & 5.790 & 5.960 & 6.132 & 6.417 & 7.509 & 7.769  \\
0.8 & 1.194 & 3.129 & 2.620 & 6.035 & 5.901 & 5.844 & 5.905 & 6.055 & 6.211 & 6.475 & 7.525 & 7.861  \\
0.9 & 1.193 & 3.156 & 2.646 & 6.259 & 6.047 & 5.946 & 5.983 & 6.119 & 6.266 & 6.514 & 7.529 & 7.937  \\
1   & 1.192 & 3.178 & 2.667 & 6.432 & 6.146 & 6.009 & 6.030 & 6.159 & 6.300 & 6.538 & 7.523 & 8.000 \\
1.5 & 1.188 & 3.236 & 2.723 & 6.635 & 6.142 & 5.966 & 5.999 & 6.135 & 6.273 & 6.493 & 7.398 & 8.169  \\
2   & 1.187 & 3.235 & 2.727 & 6.192 & 5.752 & 5.661 & 5.760 & 5.929 & 6.078 & 6.302 & 7.175 & 8.180  \\
3   & 1.185 & 3.139 & 2.649 & 4.957 & 4.825 & 4.930 & 5.143 & 5.365 & 5.533 & 5.767 & 6.617 & 7.947  \\
4   & 1.184 & 2.974 & 2.512 & 3.955 & 4.051 & 4.274 & 4.547 & 4.792 & 4.969 & 5.206 & 6.050 & 7.535  \\
5   & 1.183 & 2.777 & 2.347 & 3.226 & 3.446 & 3.726 & 4.025 & 4.279 & 4.457 & 4.694 & 5.546 & 7.040  \\
6   & 1.183 & 2.569 & 2.172 & 2.688 & 2.970 & 3.275 & 3.582 & 3.837 & 4.015 & 4.251 & 5.114 & 6.515  \\
7   & 1.183 & 2.361 & 1.996 & 2.281 & 2.589 & 2.901 & 3.207 & 3.460 & 3.636 & 3.869 & 4.727 & 5.987  \\
8   & 1.183 & 2.157 & 1.823 & 1.964 & 2.281 & 2.592 & 2.894 & 3.142 & 3.314 & 3.543 & 4.379 & 5.470  \\
9   & 1.183 & 1.962 & 1.659 & 1.707 & 2.018 & 2.314 & 2.598 & 2.825 & 2.981 & 3.182 & 3.859 & 4.976  \\
10  & 1.182 & 1.778 & 1.503 & 1.512 & 1.824 & 2.123 & 2.410 & 2.643 & 2.803 & 3.012 & 3.729 & 4.510  \\
\hline
\end{tabular}
\end{table*}

The two coefficients ($K_{V,\,\rm red}(n)$ and $3f(n)$ from Equation~\ref{eq:mass_Re_exact-2}) 
enter dynamical mass estimates of the same form, 
$M (<r_{-3}) = C(n)\, \sigma^2 R_{\rm e}/G$, but differ profoundly in their 
behaviour: $3f(n)$ varies only by $\sim$5 per cent across $n = 1.0$--$10$ 
(from 4.53 to 4.77), while $K_{V,\,\rm red}(n)$ decreases by a factor of $\sim$5 
(from $\sim$4.5 at $n=1$ to $\sim$0.9 at $n=10$).  

Notably, the ratio $3f(n)/K_{V,\,\rm red}(n)$ is not monotonic near $n\approx1$:
it dips below unity across a narrow interval, $1.05 \lesssim n \lesssim 1.46$,
reaching a minimum of $\approx0.989$ near $n\approx1.3$, before the more
typical ordering ($3f > K_{V,\,\rm red}$) resumes for larger $n$.

This striking contrast has two origins.
First, the velocity dispersions entering the two estimates differ: 
$\langle\sigma_{\rm los}^2\rangle$ is the global luminosity-weighted 
average, while $\langle\sigma_{\rm los}^2\rangle_{R_{\rm e}/10}$ samples only the 
innermost region of the galaxy. For S\'ersic indices $n \ge 1.0$, the 
internal radial velocity dispersion profile drops to zero at the origin, 
whereas for the low-index regime ($n < 1.0$) it remains finite and non-zero 
\citep{2019A&A...626A.110B}. This means that although the local dispersion 
at the origin is finite for low-index systems, the luminosity-weighted, 
aperture velocity dispersion profiles still become progressively more 
centrally peaked as $n$ increases (see, e.g., Figure~8 of 
\citealt{1997MNRAS.287..221G} and \citealt{2004RMxAA..40...69S}), partly explaining the strong reduction in 
$K_{V}(n)$ and $K_{V,\,\rm red}(n)$ with increasing $n$.

Second, the estimators differ fundamentally in their underlying physical 
assumptions regarding dark matter on larger scales. The $K_{V}(n)$ and thus 
$K_{V,\,\rm red}(n)$ coefficients are derived under a strict 
mass-follows-light assumption out to infinity, where the absence of an 
extended dark matter halo causes both the mass density and the stellar 
velocity dispersion profile to decline steeply at large radii. This steep 
outer fall-off means that the central dispersion $\sigma_0^2$ is also  
elevated relative to the global average dispersion (a discrepancy that 
worsens as $n$ increases), forcing the $K_{V,\,\rm red}(n)$ coefficient to decrease strongly (reaching
values as low as $\sim0.9$ at high $n$). 
By contrast, the $M_{-3}$ estimator is designed to remain valid when a
dominant dark-matter halo maintains an approximately flat (isothermal)\footnote{In this context, an approximately ``isothermal'' potential 
implies that the internal, total (stellar plus dark matter) mass density profile 
behaves as $\rho_{\rm tot}(r) \propto r^{-2}$ over the radial range under 
consideration, corresponding to a logarithmic potential $\phi = v_{\rm c}^2 
\log r + \text{const}$ and a linear mass profile $M_{\rm tot}(r) \propto r$. 
Under spherical symmetry and velocity isotropy, this potential keeps the 
internal 3D velocity dispersion of the tracer stars spatially constant 
($\sigma_r = v_{\rm c}/\sqrt{3}$), while the projected line-of-sight velocity 
dispersion profile remains approximately flat across the main body of the 
galaxy \citep[see, e.g.,][]{2010MNRAS.404.1165C}.}
potential. In that situation, the outer velocity-dispersion profile does not
decline sharply, and the mass inside the dynamically special radius
$r_{-3}$ stays robustly coupled to the observed dispersion, keeping
$3f(n)$ nearly constant.

Conversely, if a mass estimator that assumes a flat velocity-dispersion 
profile (such as Equations~\ref{eq:mass_R5} or \ref{eq:mass_Re_exact-2}) 
is applied to a galaxy in which mass strictly follows light, the steep 
outer decline of the stellar density will cause the luminosity-weighted 
velocity-dispersion profile to fall at large radii. This decline would violate 
the flat-profile assumption in the $M_{1/2}$ and $M_{-3}$ mass estimators and 
systematically overestimate the mass enclosed within $r_{-3}$.  
Consequently, these estimators provide a conservative upper bound: 
if they already indicate a negligible dark-matter fraction inside 
$r_{-3}$, the true enclosed mass must be even smaller, rendering the 
conclusion of a dark-matter-poor (or dark-matter-free) central region 
more secure.
For example, if the flat-dispersion (Wolf-type)
estimator is applied to an isotropic, self-gravitating Gaussian ($n=0.5$)
stellar system without dark matter, the resulting value recovers only
$\approx80$ per cent of the true mass inside $r_{-3}$. This discrepancy does
not represent a failure of the $r_{-3}$ scale itself, but rather reflects the
breakdown of the assumed flat velocity-dispersion profile: in a pure
mass-follows-light system the absence of an extended halo causes the dispersion
profile to decline steeply at large radii. For such systems, the self-consistent
virial coefficients just discussed, or the $K_V(n,R_{\rm ap})$ presented in 
Section~\ref{Sec_GC97_ext}, should be employed instead.

\subsubsection{New virial coefficients for dark-matter-free stellar systems}
\label{Sec_GC97_ext}

The aperture dependence of $K_V(n)$ reveals a genuine feature of the 
models: at fixed $n$, $K_V$ is generally not monotonic with aperture size, 
but reaches a minimum at an intermediate aperture (typically between 
$R_5/5$ and $R_5/2$ for $1\lesssim n\lesssim3$; 
Table~\ref{tab:KV_apertures}) before rising toward the infinite-aperture 
limit. This behaviour is consistent with the known radial variation of the 
projected velocity-dispersion profile of isotropic $R^{1/n}$ models 
\citep{1997A&A...321..111P}, in which the line-of-sight dispersion can 
peak away from the centre. As the aperture grows to encompass, and then 
extend beyond, that peak, the luminosity-weighted 
$\langle\sigma_{\rm los}^2\rangle$ first rises and then falls, producing 
the corresponding minimum in
\begin{equation}
K_V(n,R_{\rm ap})\;\equiv\;
\frac{G M_{\rm total}}{R_5\langle\sigma_{\rm los}^2\rangle_{R_{\rm ap}}}.
\end{equation}

The ratio $R_g/R_5$ is, likewise, non-monotonic with $n$, reaching a maximum 
of $R_g/R_5\approx2.730$ near $n\approx1.8$ (Table~\ref{tab:KV_apertures}) 
before declining at larger $n$. Over the range $0.5\le n\le10$, it is 
accurately approximated by
\begin{equation}
\frac{R_g}{R_5} \approx
3.348 - 0.211\,n + 0.00255\,n^2 - 1.120\,\mathrm{e}^{-0.845\,n}
\label{eq:Rg_R5_fit}
\end{equation}
with an accuracy of $\lesssim$0.3 per cent with the tabulated 
values. In the infinite-aperture limit, the luminosity-weighted mean-square 
velocity is related to the potential energy by the scalar virial 
theorem. For an isotropic system this immediately yields the exact 
identity
\begin{equation}
K_V(n,\infty)\;=\;3\,\frac{R_g}{R_5},
\label{eq:KV_inf_exact}
\end{equation}
so that the infinite-aperture virial coefficient is simply three times 
the ratio of the gravitational radius to the scale radius $R_5$. 
The maximum in $R_g/R_5$ therefore corresponds directly to a maximum in the
infinite-aperture virial coefficient as a function of $n$. This relation
does not, however, determine the finite-aperture coefficients, which also
depend on the aperture-dependent luminosity-weighted velocity dispersion.
For convenience, the infinite-aperture coefficient may be approximated over
the same interval by
\begin{equation}
K_V(n,\infty)\;\approx\;10.044 - 0.632\,n + 0.00765\,n^2 - 3.360\,\mathrm{e}^{-0.845\,n}
\label{eq:KV_inf_fit}
\end{equation}

\subsection{\texorpdfstring{$R_{-2}$ and $r_{-3}$ rather than $R_{\rm e}$ and $r_{1/2}$}{R-2 and r-3 rather than Re and r1/2}}
\label{Sec_illusion}

A natural question is whether a mass enclosed within a sphere of radius $r_{-3}$---which 
contains different fractions of the total light for galaxies with different 
S\'ersic indices (Table~\ref{tab:sersic_slopes})---introduces 
a systematic trend when comparing galaxies across the ETG, and thus S\'ersic, sequence. This 
concern, however, applies with equal or greater force to the conventional 
use of $r_{1/2}$. The radius enclosing 50 per cent of the \textit{stellar} 
light does not enclose 50 per cent of the \textit{total} (stellar plus dark) 
mass, with the fraction of total mass within $r_{1/2}$ varying from galaxy to 
galaxy depending on the dark matter fraction and its radial distribution 
--- quantities that may differ systematically with galaxy luminosity, environment, 
and formation history. The special significance often attributed to $r_{1/2}$ 
is therefore somewhat illusory: it is a fixed fraction of the \textit{light}, 
not of anything dynamically fundamental.

More broadly, \citet{2019PASA...36...35G} demonstrated that galaxy scaling 
relations involving $R_z$ (the radius enclosing a fraction $z$ of the total 
light) change their slope, curvature, and scatter continuously as $z$ is 
varied, with no particular value of $z$ being physically preferred. Selecting 
$z = 0.5$ (i.e., $R_e$) was an arbitrary historical choice, not a 
physically motivated one. 
While outer percentage scales such as $z = 0.8$ 
\citep[$R_{80}$; e.g.][]{2019ApJ...872L..13M} 
have been suggested to connect 
more closely with halo mass, they can probe the faint outer envelope of a galaxy 
(for $n = 4$, $R_{80} \approx 3.17\,R_{\rm e}$; for $n = 1$, $R_{80} \approx 1.78\,R_{\rm e}$)   
and therefore remain observationally vulnerable to background sky-subtraction uncertainties 
and profile truncation.
By contrast, $R_{-2}$ is defined by 
the \textit{local gradient} of the luminosity profile---a gradient-defined 
landmark that can be identified non-parametrically from the observed profile 
(Section~\ref{Sec_Non-par}) in the high signal-to-noise regime of the galaxy's 
main body, and is stable under measurement errors 
(Section~\ref{Sec_stable})---and $r_{-3}$ coincides with the dynamically special radius 
where the mass-anisotropy degeneracy in the Jeans equation is minimized \citep{2010MNRAS.404.1165C, 
2010MNRAS.406.1220W}. 
That $r_{-3}$ encloses $\sim$53--61 per cent of the total light 
(Table~\ref{tab:sersic_slopes}), varying modestly with $n$, is a 
consequence of the mathematical properties of the S\'ersic family rather than 
a design requirement---and the modest variation is no more problematic than 
the analogous variation in the fraction of total \textit{mass} enclosed within 
$r_{1/2}$ for galaxies with different dark matter fractions.
By breaking away from constant-percentage 
light boundaries and mapping the directly observable $R_5$ to the true 
$r_{-3}$ radius, the resulting mass estimator (Equation~\ref{eq:mass_R5}) remains dynamically robust 
and removes the specific systematic discrepancy introduced by substituting $r_{1/2}$ for $r_{-3}$. 
Because $R_5/R_{\rm e}$ varies only from $\approx1.20$ to $\approx1.18$ 
across the observed range of S\'ersic indices, the logarithmic size coordinate 
is shifted by a nearly constant $\approx+0.075$ dex; the curved topology of 
size--mass relations is therefore essentially unchanged. The advantage of 
$R_5$ lies in its gradient definition and its mapping onto $r_{-3}$, 
not in a reshaping of the empirical size sequence.

\subsection{Implications for ultra-diffuse galaxies}
\label{Sec_UDGs}

UDGs are not a new morphological class. 
They were previously designated ``IC~3475 type'' galaxies 
\citep{1956AJ.....61...69R, 1984AJ.....89..919S, 1985AJ.....90.1681B}, 
a population of large, low-surface-brightness systems shown to form the 
natural faint-end continuation of the dwarf early-type sequence 
\citep{2025PASA...42..155G}.  
This continuity is itself the extension 
of a broader result, in which  the 
long-alleged divide between dwarf and ordinary ETGs at an absolute $B$-band magnitude $\mathfrak{M}_B 
\approx -18$ mag was shown to be an artefact of the arbitrary percentage of light 
used to define $R_{\rm e}$ \citep{2019PASA...36...35G}.  Rather than 
evidence for distinct formation physics, dwarf and ordinary ETGs 
instead form a single, structurally non-homologous sequence traced by the  
S\'ersic index. 
In the size--luminosity and size--surface brightness diagrams, UDGs   
extend the curved relations already defined by brighter ETGs, and their 
apparently large $R_{\rm e}$ values are expected from their  
low S\'ersic indices and faint central surface brightnesses that 
characterize this regime.
Crucially, \citet{2025PASA...42..155G} revealed that, at fixed $\mathfrak{M}_B$, 
the intrinsic scatter in central surface brightness and 
S\'ersic index about the fundamental $\mathfrak{M}_B$--$\mu_{0}$ and $\mathfrak{M}_B$--$n$ 
relations propagate, through the $R^{1/n}$ model, into a corresponding 
scatter in $R_{\rm e}$ at that magnitude.  Now, because dynamical mass is often 
estimated as $\sigma^2 R_{\rm e}/G$, this structurally-driven scatter in 
$R_{\rm e}$ becomes entangled with, and can be mistaken for, genuine $n$-dependent and 
mass-dependent trends --- underscoring the need for the kind of 
non-homology-aware mass estimator developed in this paper. A companion 
paper (in preparation) will extend this picture using 
measured velocity dispersions and the $(M_{\rm dyn}/L)_{-3}$ estimator introduced 
here, to directly quantify dynamical mass-to-light trends, and inferred dark matter fractions, across the 
dwarf-to-UDG regime.
The forthcoming Legacy Survey of Space and Time 
\citep[LSST;][]{2019ApJ...873..111I} is expected to increase the known 
population of such low-surface-brightness systems by orders of magnitude, 
further underscoring the need for size and mass estimators that remain 
robust across the full range of S\'ersic index.

The relevance of the $R_5$--$r_{-3}$ connection is particularly
pronounced for low-$n$ systems.  At the 3D stellar
half-light radius $r_{1/2}$, the logarithmic slope of the deprojected
luminosity density departs increasingly from the kinematically
preferred value of $-3$ as $n$ decreases.  For $n\lesssim1$, typical of
dwarf and UDGs, the slope at $r_{1/2}$ is only
$\sim-2.4$--$-2.5$, whereas $r_{-3}$ is, by definition, the radius at
which the slope is exactly $-3$ (Table~\ref{tab:sersic_slopes}).
Consequently, identifying $r_{1/2}$ with $r_{-3}$ is a substantially
poorer approximation for these systems than for higher-$n$ galaxies,
with direct consequences for minimizing orbital anisotropy bias, and improving dynamical mass and mass-to-light-ratio
estimates.

A structural connection between UDGs and the discs of low-surface-brightness 
spiral galaxies is also known \citep[see Fig.~12 of][]{2025PASA...42..155G}. 
Indeed, direct imaging has revealed faint 
spiral arms embedded within low-luminosity, low-$n$ early-type dwarf 
galaxies \citep[e.g.,][]{2000A&A...358..845J, 2002A&A...391..823B, 2003AJ....126.1787G}, reinforcing 
the view that at least some members of this population contain an 
underlying disc component rather than being spherical, purely pressure-supported 
spheroids.
More broadly, this structural continuity is one facet of a wider 
evolutionary picture. Under the `Triangal' framework 
\citep{2023MNRAS.522.3588G, 2026MNRAS.549ag782G}, low-mass, 
low-concentration systems are linked to their more massive counterparts 
through accretion- and merger-driven pathways that naturally accommodate 
the S\'ersicification of dwarf galaxies \citep{2024MNRAS.535..299G} and 
the persistence of residual disc-like features or partial rotational 
support in the UDG regime.

The discovery of a growing class of low-luminosity, UDGs 
exhibiting unusually low stellar velocity dispersions---such as 
NGC~1052-DF2, DF4, and DF9 in the NGC~1052 group \citep{2018Natur.555..629V, 
2019ApJ...874L...5V, 2022Natur.605..435V, 2023ApJ...957....6S, 
2026ApJ..1004..210K} and, to a lesser extent, FCC~224 and FCC~240 in the outskirts of the Fornax 
cluster \citep{2026ApJ..1005...15B}---presents a unique laboratory for 
dynamical mass estimators such as those presented in this work. Characterized by low 
S\'ersic indices $n \approx 0.55$--$0.80$ \citep[][their Table~1]{2026ApJ..1005...15B}, 
their internal dynamics are consistent with being dominated by baryons alone, 
with little or no dark matter within their stellar bodies.
While some low-$n$ ETGs are flattened by tangential orbital anisotropy 
($\beta<0$, i.e.\ little net rotation), because $M_{-3}$ is evaluated at the anisotropy-insensitive 
radius $r_{-3}$, the enclosed-mass estimate retains reduced sensitivity to orbital anisotropy
near $r_{-3}$. Thus, $M_{-3}$ continues to supply a 
physically meaningful (and typically conservative) upper bound for these 
systems.

However, applying mass estimators with inherent assumed-to-be-flat, aperture velocity dispersion profiles to allegedly 
dark-matter-free systems introduces significant, albeit conservative,  
caveats that have at times been overlooked. As noted in 
Section~\ref{Sec_prior_mass}, if a low-$n$ galaxy is free of dark matter, 
then the steep outer decline of its 
S\'ersic stellar density profile will cause the luminosity-weighted aperture velocity 
dispersion profile to fall at large radii, violating the flat-profile 
assumption of both the \citet{2010MNRAS.406.1220W} mass estimator ($M_{1/2}$) and 
the refined $M_{-3}$ mass estimator (Equation~\ref{eq:mass_R5}). When this decline is 
neglected, the aperture-averaged velocity dispersion measured within 
say $R_{\rm e}$ will systematically overestimate the global average 
dispersion, causing both estimators to overpredict the actual enclosed mass. 
Fortuitously, this overlooked overprediction worked in favor of those earlier 
studies by acting as a highly conservative upper limit: if the flat-profile 
estimators already indicate a negligible dark matter fraction inside the scale 
radius, the true spatial mass must be even smaller, rendering the conclusion 
of a dark-matter-free state even more secure.

Additionally, as quantified by Equation~\ref{eq:Q_n_ratio}, the deprojected 
mass-to-light ratio inside $r_{-3}$ is systematically smaller than the value 
obtained within $r_{1/2}$ if $n\lesssim1$. For the low-index regime, the difference 
is $\sim$7.5 per cent as $n=0.5$ (Table~\ref{tab:fn_vs_KV}). 
Consequently, the true mass-to-light ratios of these dwarf systems are even 
lower (i.e.\ more dark-matter-deficient) than those inferred from the 
half-light mass estimator of \citet{2010MNRAS.406.1220W}.

For S\'ersic profiles with $n \ge 0.5$ that appear free of dark matter, it may be 
preferable to compare with an isotropic, mass-follows-light mass estimator (Section~\ref{Sec_GC97_ext}).

\subsubsection{A worked example: NGC~1052--DF2}
\label{Sec_DF2_example}

Adopting a distance of $D\approx20$\,Mpc \citep{2018Natur.555..629V,2021ApJ...914L..12S}, 
the UDG NGC~1052--DF2 has a circularized effective radius $R_{\rm e}\approx2.03$\,kpc 
(derived from $R_{\rm e,maj}=22.6''\approx2.2$\,kpc with $b/a=0.85$; and 
$R_5=(2n/b_n)^n R_{\rm e}\approx1.20\,R_{\rm e}\approx2.4$\,kpc), 
a S\'ersic index $n\approx0.60$, and a total 
$V$-band luminosity $L_V\approx1.1\times10^8\,{\rm L}_\odot$, corresponding 
to a total stellar mass $M_*\approx(1.8\text{--}2.2)\times10^8\,{\rm M}_\odot$ for 
$M_*/L_V\approx1.6$--$2.0$ \citep{2018Natur.555..629V,2022ApJ...940L...9V}. 
The lower stellar mass and $M_*/L_V$ ratio would be $\sim$20 per cent smaller ($M_*\approx1.4\times10^8\,{\rm M}_\odot$) had the authors adopted the \citet{2002Sci...295...82K} initial mass function (IMF) rather than that from \citet{2001MNRAS.322..231K}. 

Kinematics have been measured via two distinct tracer populations: 
\begin{enumerate}
  \item \textit{Diffuse stellar absorption within $\sim 1\,R_{\rm e}$}: KCWI spectroscopy yielded 
  $\sigma_{\rm los}=8.5^{+2.3}_{-3.1}\,{\rm km\,s^{-1}}$ \citep{2019ApJ...874L..12D}, 
  which adjusts to $\sigma_{\rm los}=6.3^{+3.7}_{-3.1}\,{\rm km\,s^{-1}}$ after accounting 
  for stellar macroturbulence and binary motion \citep{2026ApJ..1004..210K}; and 
  complementary MUSE spectroscopy within $1\,R_{\rm e}$ yielded 
  $\sigma_*(R_{\rm e})=10.8^{+3.2}_{-4.0}\,{\rm km\,s^{-1}}$ (adjusted to 
  $\approx9.2^{+4.5}_{-3.8}\,{\rm km\,s^{-1}}$) alongside evidence for a mild velocity 
  gradient interpreted as prolate rotation \citep{2019A&A...625A..76E,2026ApJ..1004..210K}.
  \item \textit{Globular cluster tracers}: ten compact, luminous star clusters 
  extending out to $R_{\rm out}\approx7.6\,{\rm kpc}\approx3.1\,R_5$ exhibit an 
  observed biweight velocity dispersion $\sigma_{\rm obs}\approx8.4\,{\rm km\,s^{-1}}$ 
  and an intrinsic dispersion $\sigma_{\rm intr}=3.2^{+5.5}_{-3.2}\,{\rm km\,s^{-1}}$ 
  (with $\sigma_{\rm intr}<10.5\,{\rm km\,s^{-1}}$ at 90~per~cent confidence; 
  \citealt{2018Natur.555..629V}).
\end{enumerate}

The published half-light dynamical mass and mass-to-light ratio, based on $\sigma_{\rm los}=8.5^{+2.3}_{-3.1}\,{\rm km\,s^{-1}}$, are 
$M_{1/2}=(1.3\pm0.8)\times10^8\,{\rm M}_\odot$ and 
$(M_{\rm dyn}/L_V)_{1/2}\approx2.5$ \citep{2019ApJ...874L..12D}. 
For a \citet{2002Sci...295...82K} IMF, this implies that the contribution from dark matter within the sphere of radius $r_{1/2}$ is nearly equal to that of the stellar mass, while assuming a typical old, metal-poor globular cluster mass-to-light ratio of 2 \citep{2005ApJS..161..304M}, dark matter would contribute just $\sim$15 per cent of the mass inside this sphere. 

From Table~\ref{tab:fn_vs_KV}, $Q(0.6)\equiv(M_{\rm dyn}/L)_{-3}/(M_{\rm dyn}/L)_{1/2}\approx0.941$, 
and therefore the spatial mass-to-light ratio evaluated at the 
anisotropy-insensitive radius $r_{-3}$ is $\approx6$~per~cent lower, that is, slightly more dark-matter deficient 
and leaving less room for a dark-matter contribution within $r_{-3}$. 
However, the estimators' inherent assumption of a flat velocity-dispersion profile 
due to the presence of an extended dark matter halo becomes problematic when the inferred dynamical mass approaches the suspected stellar mass. 
The estimated mass and mass-to-light ratio within a sphere equal to the scale radius may be an upper limit due to the potential existence of a declining aperture velocity dispersion profile. 
A self-consistent mass-follows-light model may, therefore, provide a more appropriate 
framework for evaluating the mass and mass-to-light ratio of this system. 

If NGC~1052--DF2 is predominantly self-gravitating in stars, the appropriate 
virial coefficient for the diffuse stellar component depends on the aperture over 
which its velocity dispersion is measured (Table~\ref{tab:KV_apertures}): 
\begin{itemize}
  \item For the \textit{diffuse stellar light} sampled within 
  $R_{\rm ap}\approx R_{\rm e}\approx4R_5/5$, $K_V(0.6,4R_5/5)\approx6.02$, yielding a total dynamical mass
  \begin{equation}
  M_{\rm tot}=K_V(n,4R_5/5)\,G^{-1}\,R_5\,\langle\sigma_{\rm los}^2\rangle_{R_{\rm e}}
  \approx(1.33\text{--}2.43)\times10^8\,{\rm M}_\odot
  \end{equation}
  for $\sigma_{\rm los}=6.3$--$8.5\,{\rm km\,s^{-1}}$ (corresponding to global $(M_{\rm dyn}/L_V)_{\rm tot}\approx1.2$--$2.2$). 
  The corrected MUSE value of $9.2\,{\rm km\,s^{-1}}$ yields $M_{\rm tot}\approx2.84\times10^8\,{\rm M}_\odot$ and $(M_{\rm dyn}/L_V)_{\rm tot}\approx 2.6$.
Given the plausible range of stellar mass-to-light ratios from $\sim$1.4--2.0, this latter result suggests NGC~1052–DF2 could contain some dark matter.
  \item For comparison, the \textit{globular cluster system} provides an outer kinematic tracer extending to $R_{\rm out}\approx3.1\,R_5$. 
  If its raw observed dispersion ($\sigma_{\rm obs}\approx8.4\,{\rm km\,s^{-1}}$) is provisionally compared with the infinite-aperture coefficient of the mass-follows-light $R^{1/n}$ model, 
  $K_V(0.6,\infty)\approx7.66$, one obtains 
  \begin{equation}
  M_{\rm tot}\approx K_V(n,\infty)\,G^{-1}\,R_5\,\sigma_{\rm obs}^2\approx3.02\times10^8\,{\rm M}_\odot,
  \end{equation}
  yielding $(M_{\rm dyn}/L_V)_{\rm tot}\approx2.7$, with $M_{\rm tot}<4.71\times10^8\,{\rm M}_\odot$ and 
  $(M_{\rm dyn}/L_V)_{\rm tot}\lesssim4.3$ at 90~per~cent confidence for $\sigma_{\rm intr}<10.5\,{\rm km\,s^{-1}}$.
However, because $\sigma_{\rm obs}$ includes observational measurement errors of $\sim 7$--$8\,{\rm km\,s^{-1}}$, 
  evaluating the mass estimator with the best-fit intrinsic dispersion ($\sigma_{\rm intr}=3.2^{+5.5}_{-3.2}\,{\rm km\,s^{-1}}$) 
  instead yields $M_{\rm tot}\approx0.44\times10^8\,{\rm M}_\odot$ ($(M_{\rm dyn}/L_V)_{\rm tot}\approx0.4$). 
While this `central-point' estimate is lower than expected for standard stellar populations, the large upper uncertainty 
  on the intrinsic dispersion ($\sigma \lesssim 8.7\,{\rm km\,s^{-1}}$ at $1\sigma$) comfortably encompasses the 
  expected stellar mass scale ($M_*\approx(1.4\text{--}2.0)\times10^8\,{\rm M}_\odot$). Reducing the uncertainty on the recessional velocities of the globular clusters would be helpful.
\end{itemize}  

Taken together, the stellar-absorption measurements and the globular-cluster kinematics place the dynamical mass of NGC~1052--DF2 on a scale broadly consistent with its stellar mass, rather than requiring a massive dark-matter halo. 
A broader application of these estimators across a larger ensemble of UDGs will be presented in a companion paper, providing the dynamical counterpart to the structural analysis of UDGs given in \citet{2025PASA...42..155G}.

\section{Summary and Conclusions}
\label{Sec_Conclusions}

$R_{\rm e}$, or not $R_{\rm e}$, that is the question explored here. 
The projected half-light radius \(R_{\rm e}\) has been the default size metric for galaxies for more than half a century. While convenient, it is an arbitrary 50-per-cent light boundary that mixes structural non-homology into every scaling relation that employs it. 
As discussed in Section~\ref{Sec_illusion}, in most galaxies, the radius that contains 50 per cent of the stellar light is not expected to correspond to the radius enclosing 50 per cent of the total mass (stellar plus dark matter). Consequently, the commonly ascribed importance of half-light radii, including the 3D counterpart $r_{1/2}$, is somewhat misleading: it simply marks a fixed fraction of the light, rather than representing any dynamically meaningful quantity.
This paper replaces that arbitrary scale with a physically motivated alternative with observational advantages. 

The projected radius \(R_{5} \equiv R_{-2}\), at which the local logarithmic slope of the surface brightness profile equals $5$---and the local logarithmic slope of the intensity profile equals $-2$---is introduced.  For the S\'ersic \(R^{1/n}\) family, the analytical solution is 
\begin{equation}
R_{5} \equiv R_{-2} = \left(\frac{2n}{b_n}\right)^n R_{\rm e}, \nonumber 
\end{equation}
which converges to the elegant limit \(e^{1/6}\approx 1.181\,R_{\rm e}\) as \(n\to\infty\) (Section~\ref{Sec_linear}). 
This formulation also maps directly onto traditional disc scaling parameters; 
for an exponential model ($n=1$), the $R_5$ scale radius reduces to $R_5 = 2h$, completely bypassing the $b_n$ parameter.

By adopting \(R_{5}\) (and the associated intensity \(I_{5}\)) as the fundamental scale, a completely $b_n$-free reparameterization of the $R^{1/n}$ light profile model is developed, such that 
\begin{equation}
I(R) = I_0 \exp \left[ -2n \left(\frac{R}{R_{5}}\right)^{1/n} \right]
= 
I_{5} \exp \left\{ -2n \left[ \left(\frac{R}{R_{5}}\right)^{1/n} - 1 \right] \right\}. \nonumber
\end{equation}
The re-expression for the S\'ersic function now depends only on \(R_5 \equiv R_{-2}\), \(n\), and a normalization intensity, 
bypassing the coupling parameter $b_n$.  
Furthermore, because $\mu(R) \propto 5\log R$ at/near $R_5$, measurement errors in \(R_5\) and \(\mu_5\ \equiv -2.5\log I_5\) now have the desirable property that they largely cancel so that the total luminosity and magnitude ($\propto \mu_5 -5\log R_5$) remain robust, offering a stability not present with the measurement of $R_{\rm e}$ and $\mu_{\rm e}$.  
Real data with noise, PSF effects, and departures from a S\'ersic profile remain to be tested, but are expected to undermine the robustness expected from the analytical models presented herein. 
The \citet{1997A&A...321..111P} model that approximates the deprojection of the \citet{1963BAAA....6...41S} $R^{1/n}$ model has also been re-expressed using $I_5$ and $R_5$ (Section~\ref{Sec_PS}). 

The scale radius \(R_5\) can also be measured non-parametrically: it is simply 
the radius at which the local logarithmic slope of the observed surface-brightness 
profile equals 5\,mag\,dex\(^{-1}\). Once \(R_5\) is located, a concentration 
index formed from the light interior to \(R_5/3\) and \(R_5\) uniquely 
determines both the S\'ersic index and the total-luminosity extrapolation 
factor, yielding a fitting-free estimate of the total magnitude 
(Section~\ref{Sec_Non-par-2}).
Furthermore, the asymptotic Kron radius $R_1(\infty)$ is exactly 
equal to $R_5$ when $n=1$ (Equation~\ref{eq:R1_R5_identity}), providing a physical bridge between 
non-parametric apertures and the $R_5$ scale radius.

The expressions herein improve and unify the 2D and 3D  
schemes of \citet{2010MNRAS.404.1165C} and \citet{2010MNRAS.406.1220W} 
across the full range of observed S\'ersic indices observed in galaxies. 
Specifically, the projected local sweet spot $R_{-2}$ and the intrinsic spatial sweet spot $r_{-3}$ are shown to be different dimensional manifestations of the same underlying physical transition, decoupled into separate spatial scales by the profile curvature of S\'ersic systems. 
By replacing those works' respective 
power-law and spatial half-light approximations, a refined, $n$-dependent mass estimator $M_{-3}$ and 
spatial mass-to-light ratio $(M_{\rm dyn}/L)_{-3}$ are derived across 
all structural regimes ($n = 0.25\text{--}10$).
The \((M_{\rm dyn}/L)_{-3}\) presented here is more robust to orbital anisotropy, and systematically more conservative for $n<1$ systems, than the half-light mass estimator \((M_{\rm dyn}/L)_{1/2}\) of \citet{2010MNRAS.406.1220W}, with significance for studies of dwarf ETGS, including UDGs.

While \citet{2010MNRAS.406.1220W} documented that the spatial radius 
$r_{-3}$ is systematically larger than $r_{1/2}$ (tabulating these 
ratios in their Appendix B), they did not evaluate how the 
associated deprojected luminosity and resulting mass-to-light ratio 
shift when transitioning between these scales. 
The systematic $n$-dependent  offset of \((M_{\rm dyn}/L)_{1/2}\) from the dynamically and structurally preferred \((M_{\rm dyn}/L)_{-3}\) is described by the smooth function \(Q(n)\), derived here (Equation~\ref{eq:Q_n_ratio}) and quantified in Table~\ref{tab:fn_vs_KV}. 

An exact single-integral identity for the (3D) enclosed-light 
fraction  has also been derived here. 
Given any projected aperture radius $R=r$, the identity converts the observable projected light fraction 
$F_{\rm 2D}(r)$ into the  fraction $F_{\rm 3D}(r)$ within a sphere of radius $r$, with a single one-dimensional integral (Equation~\ref{eq:F3D_single_integral}). 
It collapses the classical nested double integrals of the Abel deprojection 
into a single, numerically stable quadrature, aiding the route from projected photometry to spatial mass-to-light ratios.

Two particularly compact asymptotic results emerge in the formal limit
$n\to\infty$: the anisotropy-insensitive radius converges to
$r_{-3}/R_5\to{\rm e}/2\approx1.359$, while the spatial mass-to-light scaling
function converges to $h(n)\to{\rm e}\approx2.718$. Both limits follow from a
Laplace-method expansion of the exact Abel deprojection and provide simple,
memorable anchors for the finite-$n$ expressions tabulated in this work.

Key conversions and estimators are collected for convenience in the 
quick-reference cookbook of Appendix~\ref{Appdx_cookbook} 
(Table~\ref{tab:cookbook}) and are implemented in the public 
\textsc{R5Tools} package and web application mentioned in Section~\ref{Sec_data_avail}.

For practical applications, the procedure is straightforward: determine
$R_5$ from the observed logarithmic surface-brightness gradient or a parametrized fit to the light profile, determine
$n$ either from this $R^{1/n}$ fit or from the $R_5$-based concentration index (Section~\ref{Sec_Non-par}),
evaluate $r_{-3}=g(n)R_5$ ($\equiv f(n)R_{\rm e}$), and combine this radius with the appropriate
luminosity-weighted velocity-dispersion measurement to obtain
$M_{-3}$ and $(M_{\rm dyn}/L)_{-3}$. The corresponding (3D) 
light fraction can be evaluated directly from the single-integral identity
derived in Equation~\ref{eq:F3D_single_integral} and tabulated in Table~\ref{tab:light_fractions}. 

Taken together, these results provide a practical and theoretically cleaner framework for measuring galaxy sizes, dynamical masses, and dark-matter fractions. They motivate the use of \(R_{-2} \equiv R_5\) and \(\mu_5\) 
as complementary, and in many applications preferable, structural measures
to \(R_{\rm e}\) and \(\mu_{\rm e}\) 
(and \(M_{-3}\) and \((M_{\rm dyn}/L)_{-3}\) compared to \(M_{1/2}\) and \((M_{\rm dyn}/L)_{1/2}\)), 
in future photometric (and dynamical) analyses of ETGs. 

For systems with a measured $(M_{\rm dyn}/L)_{-3}$ that implies little or no
dark matter, this estimate may provide a conservative upper limit on the
enclosed mass, modulo net rotation and provided that the assumptions underlying the flat,
luminosity-weighted aperture velocity-dispersion profile are satisfied.  
For 
(dark matter)-free systems having light profiles with $n>0.5$, the mass-follows-light mass estimator (Sections~\ref{Sec_Bertin} and \ref{Sec_GC97_ext}) may be more applicable, as tested in Section~\ref{Sec_UDGs}.

\section*{Acknowledgements}

The author gratefully acknowledges Mike Russell, former Mathematics and Computer Science teacher at Billanook College, Victoria, Australia. 
The Author is also thankful to The University of Florida for hosting him during a 2-month sabbatical in 2019, when the foundation of this work was established and Figures~\ref{Fig0A} and \ref{Fig0B} created.
 
Publication costs were funded through the Australia and New Zealand Institutions (Council of Australian University Librarians affiliated) Open Access Agreement.
This research has used the SAO/NASA Astrophysics Data System (ADS) bibliographic services.

\section{Data Availability}
\label{Sec_data_avail}

The interactive Python package \textsc{R5Tools}, providing both analytical 
conversions and numerical routines for $R_5$-based structural, deprojection, 
and dynamical calculations, is publicly available on GitHub at 
\url{https://github.com/A-Graham/R5Tools} and is permanently archived on 
Zenodo under DOI: \href{https://doi.org/10.5281/zenodo.21992060}{10.5281/zenodo.21992060} 
\citep{Graham_R5Tools_2026}.
An interactive web interface is also provided at \url{https://r5tools.streamlit.app/}

\bibliographystyle{mnras}
\bibliography{Paper-R_2}{}

\appendix
\renewcommand{\theHtable}{A\arabic{table}}
\renewcommand{\theHfigure}{A\arabic{figure}}
\renewcommand{\theHequation}{A\arabic{equation}}

\section{Derivation of the 3D enclosed light fraction}
\label{Sec_Appdx_A}

A single-integral expression for the fraction of total light
enclosed within a sphere of radius $r$, starting from the Abel deprojection
of the S\'ersic intensity profile, is derived here. 

\subsection*{Setup}

The spatial (3D) luminosity density $\nu(r)$ is related to the
projected intensity $I(R)$ by the Abel integral equation
\citep{1991A&A...249...99C}:
\begin{equation}
\nu(r) = -\frac{1}{\pi} \int_r^\infty \frac{\mathrm{d}I/\mathrm{d}R}{\sqrt{R^2 - r^2}}\,\mathrm{d}R.
\label{eq:abel_app}
\end{equation}
The spherically enclosed 3D luminosity within radius $r$ is
\begin{equation}
L(r) = 4\pi \int_0^r \nu(r')\,r'^2\,\mathrm{d}r'.
\label{eq:L3D_def}
\end{equation}
Substituting Equation~(\ref{eq:abel_app}) into Equation~(\ref{eq:L3D_def})
gives a nested double integral:
\begin{equation}
L(r) = -4 \int_0^r r'^2 \left[
  \int_{r'}^\infty \frac{\mathrm{d}I/\mathrm{d}R}{\sqrt{R^2 - r'^2}}\,\mathrm{d}R
\right] \mathrm{d}r'.
\label{eq:L3D_double}
\end{equation}

\subsection*{Exchanging the order of integration}

The domain of integration in Equation~(\ref{eq:L3D_double}) is
$\{(r', R) : 0 \le r' \le r,\; R \ge r'\}$.
For a S\'ersic profile, the double integral is absolutely convergent,
allowing the order of integration to be exchanged by Fubini's theorem.
Splitting the outer $R$-integral at
$R = r$, the domain decomposes into two regions:
\begin{align}
\text{Region I:}  &\quad 0 \le r' \le R \le r, \label{eq:regionI}\\
\text{Region II:} &\quad 0 \le r' \le r < R < \infty. \label{eq:regionII}
\end{align}
Reversing the order of integration in each region:
\begin{equation}
\begin{split}
L(r) = -4 \biggl[ & \int_0^r \frac{\mathrm{d}I}{\mathrm{d}R}
    \underbrace{\int_0^R \frac{r'^2}{\sqrt{R^2-r'^2}}\,\mathrm{d}r'}_{\mathcal{I}_1(R)}\,\mathrm{d}R \\
  & + \int_r^\infty \frac{\mathrm{d}I}{\mathrm{d}R}
    \underbrace{\int_0^r \frac{r'^2}{\sqrt{R^2-r'^2}}\,\mathrm{d}r'}_{\mathcal{I}_2(R,r)}\,\mathrm{d}R \biggr].
\end{split}
\label{eq:L3D_swapped}
\end{equation}

\subsection*{Evaluating the inner integrals}

For $\mathcal{I}_1(R)$, the substitution $r' = R\sin\theta$ gives
\begin{equation}
\mathcal{I}_1(R) = \int_0^{\pi/2} \frac{R^2\sin^2\theta}{R\cos\theta}\,
  R\cos\theta\,\mathrm{d}\theta = R^2 \int_0^{\pi/2} \sin^2\theta\,\mathrm{d}\theta
  = \frac{\pi R^2}{4}.
\label{eq:I1}
\end{equation}

For $\mathcal{I}_2(R,r)$ with $R > r$, the same substitution
$r' = R\sin\theta$ maps $r' \in [0,r]$ to
$\theta \in [0,\arcsin(r/R)]$:
\begin{align}
\mathcal{I}_2(R,r)
&= R^2 \int_0^{\arcsin(r/R)} \sin^2\theta\,\mathrm{d}\theta \nonumber\\
&= R^2 \left[
\frac{\theta}{2} - \frac{\sin 2\theta}{4}
\right]_0^{\arcsin(r/R)} \nonumber\\
&= R^2 \left[
\frac{1}{2}\arcsin\!\left(\frac{r}{R}\right)
-\frac{r}{2R}\sqrt{1-\frac{r^2}{R^2}}
\right] \nonumber\\
&= \frac{R^2}{2}\left[
\arcsin\!\left(\frac{r}{R}\right)
-\frac{r}{R}\sqrt{1-\frac{r^2}{R^2}}
\right].
\label{eq:I2_intermediate}
\end{align}
This can be written more compactly as:
\begin{equation}
\mathcal{I}_2(R,r) = \frac{R}{2}\left[
  R\arcsin\!\left(\frac{r}{R}\right) - r\sqrt{1-\frac{r^2}{R^2}}
\right].
\label{eq:I2}
\end{equation}

\subsection*{Integration by parts on Region I}

Substituting $\mathcal{I}_1(R) = \pi R^2/4$ into the Region~I
contribution:
\begin{equation}
-4 \int_0^r \frac{\mathrm{d}I}{\mathrm{d}R} \cdot \frac{\pi R^2}{4}\,\mathrm{d}R
= -\pi \int_0^r R^2 \frac{\mathrm{d}I}{\mathrm{d}R}\,\mathrm{d}R.
\end{equation}
Integrating by parts with $u = R^2$ and $dv = (\mathrm{d}I/\mathrm{d}R)\mathrm{d}R$:
\begin{equation}
-\pi \left[ R^2 I(R) \right]_0^r + 2\pi \int_0^r I(R)\,R\,\mathrm{d}R
= -\pi r^2 I(r) + 2\pi \int_0^r I(R)\,R\,\mathrm{d}R,
\end{equation}
where $I(0)$ is finite and the $R^2$ factor ensures the boundary term
at $R=0$ vanishes.

Recognizing that $L_{\rm tot} = 2\pi \int_0^\infty I(R)\,R\,\mathrm{d}R$ (from
the projected luminosity), the 2D enclosed fraction at $r$ is
\begin{equation}
F_{\rm 2D}(r) = \frac{2\pi\int_0^r I(R)\,R\,\mathrm{d}R}{L_{\rm tot}},
\end{equation}
so the Region~I contribution to $L(r)/L_{\rm tot}$ is:
\begin{equation}
F_{\rm 2D}(r) - \frac{\pi r^2 I(r)}{L_{\rm tot}}.
\label{eq:regionI_result}
\end{equation}

\subsection*{Combining both regions}

Substituting $\mathcal{I}_2$ from Equation~(\ref{eq:I2}) into the
Region~II contribution and combining with
Equation~(\ref{eq:regionI_result}):
\begin{equation}
\begin{split}
F_{\rm 3D}(r) & = F_{\rm 2D}(r) - \frac{\pi r^2 I(r)}{L_{\rm tot}} \\
  & \quad - \frac{4}{L_{\rm tot}} \int_r^\infty \frac{\mathrm{d}I}{\mathrm{d}R} \cdot
   \frac{R}{2}\left[R\arcsin\!\left(\frac{r}{R}\right)
   - r\sqrt{1-\frac{r^2}{R^2}}\right] \mathrm{d}R.
\end{split}
\label{eq:combined_step}
\end{equation}
The Region~II integral is now integrated by parts, with
$u = \mathcal{I}_2(R,r)$
and $\mathrm{d}v = (\mathrm{d}I/\mathrm{d}R)\mathrm{d}R$. The boundary term at $R\to\infty$ vanishes
because $I(R)$ decays exponentially.
Integration by parts on the Region~II contribution, using the boundary
value $u(r)=\pi r^2/4$ together with the already-established Region~I
result, causes the terms proportional to $r^2 I(r)$ to cancel.
The derivative of $u$ is
\begin{equation}
\frac{\mathrm{d}u}{\mathrm{d}R}
=R\left[
\arcsin\!\left(\frac{r}{R}\right)
-\frac{r}{\sqrt{R^2-r^2}}
\right]
\end{equation}
Collecting the remaining terms gives the compact identity
\begin{equation}
F_{\rm 3D}(r) = F_{\rm 2D}(r) + \frac{4}{L_{\rm tot}}
  \int_r^\infty I(R)\,R\left[
    \arcsin\!\left(\frac{r}{R}\right)
    - \frac{r}{\sqrt{R^2 - r^2}}
  \right]\mathrm{d}R, 
\label{eq:F3D_final_app}
\end{equation}
which is Equation~(\ref{eq:F3D_single_integral}) of the main text.
Since $\arcsin(r/R) < \pi/2$ and
$r/\sqrt{R^2-r^2} > r/R$ for all $R > r > 0$, the bracket is strictly
negative for all $R > r$, proving $F_{\rm 3D}(r) < F_{\rm 2D}(r)$ for all
finite $r$: the spherically enclosed 3D light fraction is always less
than the projected 2D fraction at the same radius, reflecting the
contribution of foreground and background light within the projected
cylinder but outside the sphere.

\noindent\textit{Numerical verification.} Equation~(\ref{eq:F3D_final_app})
has been verified numerically against direct integration of
$4\pi\int_0^r \nu(r')r'^2\,\mathrm{d}r'$ (using the Abel-deprojected S\'ersic
profile) for $n = 1, 2, 4, 6, 8, 10$ at multiple radii, yielding
agreement to better than one part in $10^{5}$ in all cases.

\section{An exact asymptotic limit for \texorpdfstring{\lowercase{$r$}$_{-3}/R_5$}{r-3/R5}}
\label{Appdx_Sec_r3_asymptotic}

The location of $r_{-3}$ in the $n\to\infty$ limit can be obtained by
an asymptotic expansion of the exact Abel deprojection integral
(Equation~\ref{eq:exact_deproj_R5_dimensionless}). Substituting
$R=r\cosh\phi$ \citep[e.g.,][]{1990ApJ...361...78B, 2002MNRAS.333..510T}, 
so that $\mathrm{d}R/\sqrt{R^2-r^2}=\mathrm{d}\phi$) converts the
deprojection integral into
\begin{equation}
\nu(r) \propto \int_0^\infty
\exp\!\left[-2n\,{\rm e}^{t/n}\right] {\rm e}^{t/n} {\rm e}^{-t}\,\mathrm{d}\phi,
\qquad
t \equiv s+\ln(\cosh\phi),
\label{eq:coshphi_sub}
\end{equation}
where $s\equiv\ln(r/R_5)$. Expanding the exponent for large $n$ at
fixed $\phi$,
\begin{equation}
-2n\,{\rm e}^{t/n}
=
-2n-2t-\frac{t^2}{n}
+\mathcal{O}(n^{-2}),
\end{equation}
so, apart from multiplicative factors independent of $s$, the
integrand of Equation~(\ref{eq:coshphi_sub}) becomes
\begin{equation}
{\rm e}^{-2n}\,{\rm e}^{-3t}
\left[1+\frac{t-t^2}{n}+\mathcal{O}(n^{-2})\right].
\end{equation}
The leading ($n^0$) term gives $\nu(r)\propto {\rm e}^{-3s}$, showing that,
at fixed $r/R_5$, the deprojected profile approaches a power law of
slope $-3$ as $n\to\infty$, consistent with the exact deprojection of
the limiting projected $R^{-2}$ profile. The precise location of
$r_{-3}$ for large but finite $n$ is therefore fixed by the first-order
($n^{-1}$) correction.

Expanding $t-t^2$ in powers of $s$ and integrating term by term against
$\mathrm{sech}^3\phi$ gives
\begin{equation}
\int_0^\infty \mathrm{sech}^3\phi\,(t-t^2)\,\mathrm{d}\phi
=
-A\,s^2+(A-2B)\,s+C,
\end{equation}
where
\begin{equation}
\begin{aligned}
A &\equiv \int_0^\infty\mathrm{sech}^3\phi\,\mathrm{d}\phi = \frac{\pi}{4},\\
B &\equiv \int_0^\infty\ln(\cosh\phi)\,\mathrm{sech}^3\phi\,\mathrm{d}\phi
  = \frac{\pi}{4}\Bigl(\ln2-1/2\Bigr), 
\end{aligned}
\end{equation}
and the $s$-independent term $C$ does not affect the location of the
extremum. The value of $B$ follows by logarithmic differentiation of
the standard integral
\[
\int_0^\infty\mathrm{sech}^p\phi\,\mathrm{d}\phi
=
\frac{\sqrt{\pi}\,\Gamma(p/2)}
{2\,\Gamma[(p+1)/2]}
\]
with respect to $p$, evaluated at $p=3$, together with the digamma
values $\psi(2)=1-\gamma$ and
$\psi(3/2)=2-\gamma-2\ln2$.

Requiring the logarithmic density slope
$\mathrm{d}\ln\nu/\mathrm{d}s$ to equal exactly $-3$, or equivalently
requiring the first-order correction to be stationary in $s$, gives
\begin{equation}
s_\star=\frac{A-2B}{2A}.
\label{eq:sstar}
\end{equation}
Substitution of $A$ and $B$ yields the remarkably simple result
$s_\star=1-\ln2$, and therefore
\begin{equation}
\frac{r_{-3}}{R_5}
\;\longrightarrow\;
{\rm e}^{1-\ln2}
=\frac{{\rm e}}{2}
\approx1.3591
\qquad(n\to\infty).
\label{eq:r3_R5_exact}
\end{equation}

Although a pure $R^{-2}$ intensity law deprojects to a density profile
of slope $-3$ at every radius (so that $r_{-3}$ is formally undefined),
the finite-$n$ S\'ersic sequence approaches this limiting power law
through a transition region whose location relative to $R_5$ converges
to the finite value derived above.

Combined with $F_{\rm 3D}(r_{-3})\to1/2$ in the same limit
(Table~\ref{tab:light_fractions}), the composite function $h(n)$
defined in Equation~(\ref{eq:hn_definition}) has the closed-form
asymptotic limit
\begin{equation}
h(n)\;\longrightarrow\; \frac{{\rm e}/2}{1/2} = {\rm e}\approx2.7183
\qquad(n\to\infty).
\label{eq:h_asymptotic}
\end{equation}

\section{A Quick-Reference Cookbook}
\label{Appdx_cookbook}

\begin{table*}
\centering
\caption{Key $R_5$ formulas, conversions, and dynamical estimators as functions of S{\'e}rsic index $n$. For quick calculations without re-fitting profiles, existing catalog parameters ($R_{\rm e}, \mu_{\rm e}, n$) map directly to the gradient-defined framework.}
\label{tab:cookbook}
\renewcommand{\arraystretch}{1.35}
\begin{tabular}{llp{7.1cm}} 
\hline
\textbf{Physical Quantity} & \textbf{Exact Expression / Definition} & \textbf{Practical Approximation / Asymptotic Limit} \\
\hline
\multicolumn{3}{l}{\textit{\textbf{A. Structural Parameter Conversions (2D Projected)}}} \\
Scale radius ratio $R_5/R_{\rm e}$ & $R_5 = (2n/b_n)^n R_{\rm e}$ & $R_5 \approx (1.18\text{--}1.20)\,R_{\rm e}$ for $n \ge 0.5$; exact limit ${\rm e}^{1/6} \approx 1.1814$ ($n\to\infty$) \\
Disc scale-length ($n = 1$) & $R_5 = 2h = R_1(\infty)$ & Matches asymptotic Kron radius $R_1(\infty)$ exactly \\
Surface brightness at $R_5$ & $\mu_5 = \mu_{\rm e} - \frac{2.5}{\ln 10}(b_n - 2n)$ & $\Delta\mu \equiv \mu_5 - \mu_{\rm e} \approx 0.30\text{--}0.36$\,mag (rises from $0.296$ at $n=0.25$ to $0.362$ as $n\to\infty$) \\
Central surface brightness & $\mu_0 = \mu_5 - \frac{5n}{\ln 10}$ & $\mu_0 \approx \mu_5 - 2.1715\,n$ (completely $b_n$-free) \\
Total apparent magnitude & $m_{\rm tot} = \mu_5 - 5\log_{10} R_5 - 2.5\log_{10}\left[\frac{\pi {\rm e}^{2n}\Gamma(2n+1)}{(2n)^{2n}}\right]$ & Eliminates $b_n$ \\
Projected enclosed fraction & $L_{\rm 2D}(< R_5)/L_{\rm tot} = \gamma(2n, 2n)/\Gamma(2n)$ & $\approx \frac{1}{2} + \frac{1}{6\sqrt{\pi n}}$; spans $53\text{--}63\%$ for $0.5 \le n \le 10$, tending to $50\%$ as $n\to\infty$ \\
Concentration index & $\mathcal{F}_{0.33/5} \equiv L(< R_5/3)/L(< R_5) = \frac{\gamma(2n, 2n\cdot 3^{-1/n})}{\gamma(2n, 2n)}$ & Non-parametric route to $n$ and $L_{\rm tot}$ without profile fitting \\
\hline
\multicolumn{3}{l}{\textit{\textbf{B. Anisotropy-Insensitive Sweet-Spot \& Dynamical Mass (Systems with DM Halo)}}} \\
Sweet-spot radius $r_{-3}$ & $r_{-3} = g(n)\,R_5 = f(n)\,R_{\rm e}$ & For $0.5 \le n \le 10$: $g(n) \approx 1.347 - 0.097\,n^{-1} + 0.018\,n^{-2}$; exact limit $g(\infty) = {\rm e}/2 \approx 1.3591$ \\
& & For $0.5 \le n \le 10$: $f(n) \approx 1.594 - 0.104\,n^{-1} + 0.021\,n^{-2}$; exact limit $f(\infty) = \frac{1}{2}{\rm e}^{7/6} \approx 1.6060$ \\
Refined Wolf-type mass & $M_{-3} \equiv M(< r_{-3}) = 3\,G^{-1}\langle\sigma_{\rm los}^2\rangle\,g(n)\,R_5$ & $M_{-3} \approx 4.0\,G^{-1}\langle\sigma_{\rm los}^2\rangle\,R_5$ for $n \gtrsim 2$ (converging to $\approx 4.08\,G^{-1}\langle\sigma_{\rm los}^2\rangle\,R_5$ as $n\to\infty$) \\
& $\phantom{M_{-3}} = 3\,G^{-1}\langle\sigma_{\rm los}^2\rangle\,f(n)\,R_{\rm e}$ & $M_{-3} \approx 4.72\,G^{-1}\langle\sigma_{\rm los}^2\rangle\,R_{\rm e}$ for $n \gtrsim 2$ (converging to $\approx 4.82\,G^{-1}\langle\sigma_{\rm los}^2\rangle\,R_{\rm e}$ as $n\to\infty$) \\
\hline
\multicolumn{3}{l}{\textit{\textbf{C. Spatial Mass-to-Light Ratios \& Non-Homology Offsets}}} \\
Spatial mass-to-light ratio & $(M_{\rm dyn}/L)_{-3} = \frac{3}{G} h(n) \frac{R_5 \langle\sigma_{\rm los}^2\rangle}{L_{\rm tot}}$ & For $0.25 \le n \le 10$: $h(n) \approx 2.565 - 0.448\,n^{-1} + 0.066\,n^{-2}$; exact limit $h(\infty) = {\rm e} \approx 2.7183$ \\
Ratio relative to half-light & $Q(n) \equiv \frac{(M_{\rm dyn}/L)_{-3}}{(M_{\rm dyn}/L)_{1/2}} = \left(\frac{r_{-3}}{r_{1/2}}\right)\frac{0.5}{F_{\rm 3D}(r_{-3})}$ & For $0.25 \le n \le 10$: $Q(n) \approx 1.185 - 0.260\,n^{-1/2} + 0.053\,n^{-1}$; exact limit $Q(\infty) \approx 1.185$ \\
& & $Q < 1$ for $n \lesssim 1.34$ (dwarfs/UDGs), $Q > 1$ for $n \gtrsim 1.34$ (brighter ETGs) \\
\hline
\multicolumn{3}{l}{\textit{\textbf{D. Mass-Follows-Light Systems (Dark-Matter-Free / Self-Gravitating Stellar Systems)}}} \\
Total dynamical mass & $M_{\rm tot} = K_V(n, R_{\rm ap})\,G^{-1} R_5 \langle\sigma_{\rm los}^2\rangle_{R_{\rm ap}}$ & Use aperture-dependent $K_V(n, R_{\rm ap})$ defined w.r.t.\ $R_5$ (Table~5); or convert catalog $R_{\rm e}$ to $R_5$ via $R_5 = (2n/b_n)^n R_{\rm e}$ \\
Infinite-aperture limit & $K_V(n, \infty) = 3\,R_{\rm g}/R_5$ & For $0.5 \le n \le 10$: $K_V(n, \infty) \approx 10.044 - 0.632\,n + 0.00765\,n^2 - 3.360\,{\rm e}^{-0.845 n}$ \\
Central-aperture mass ($R_{\rm e}/10$) & $M_{\rm tot} = K_V(n, R_{\rm e}/10)\,G^{-1} R_{\rm e} \langle\sigma_{\rm los}^2\rangle_{R_{\rm e}/10}$ & $K_V(n, R_{\rm e}/10) \approx 3.155 + 7.221\,n - 2.633\,n^2$ for $0.5 \le n \le 1.5$ (Eq.~90) \\
& & $K_V(n, R_{\rm e}/10) \approx -0.804 + 27.35\,n^{-1} - 21.08\,n^{-2}$ for $1.5 < n \le 10$ (Eq.~90) \\
\hline
\end{tabular}
\end{table*}

\bsp    
\label{lastpage}
\end{document}